%% file: main.tex
\documentclass[11pt]{article}
\usepackage[margin=1in]{geometry} 
\usepackage[T1]{fontenc}
\usepackage[utf8]{inputenc} 
\usepackage{lmodern}
\usepackage{microtype}

\usepackage{authblk}
\usepackage{graphicx}
\usepackage{wrapfig}
\usepackage{amsmath}
\usepackage{amssymb}
\usepackage{amsthm}
\usepackage{color}
\usepackage{xurl}
\usepackage{hyperref}
\usepackage{enumitem}
\usepackage{braket}
\usepackage{enumitem}
\usepackage{booktabs}
\usepackage{stmaryrd}
\usepackage{tikz}
\usetikzlibrary{quantikz2}
\usepackage{setspace}
\usepackage[table,xcdraw]{xcolor}
\usepackage{bm}
\usepackage{float}
\usepackage{placeins}

\usepackage{tabularx}
\usepackage{booktabs}

\usepackage{caption}
\usepackage[labelformat=simple]{subcaption}

\usepackage{circuitikz}
\usetikzlibrary{shapes.misc}
\usetikzlibrary{shapes.geometric}
\usetikzlibrary{decorations.pathmorphing}

\usepackage{algpseudocode}
\usepackage[linesnumbered,ruled,vlined]{algorithm2e}

\newcommand{\be}{\begin{equation}}
\newcommand{\ee}{\end{equation}}
\newcommand{\ba}{\begin{array}}
\newcommand{\ea}{\end{array}}
\newcommand{\bea}{\begin{eqnarray}}
\newcommand{\eea}{\end{eqnarray}}

\newcommand{\calB}{{\cal B }}

\newcommand{\EE}{\mathbb{E}}

\newcommand{\app}[1]{\hyperref[app:#1]{Appendix~\ref*{app:#1}}}

\renewcommand{\sec}[1]{\hyperref[sec:#1]{Section~\ref*{sec:#1}}}
\newcommand{\ssec}[1]{\hyperref[ssec:#1]{Subsection~\ref*{ssec:#1}}}
\newcommand{\fig}[1]{\hyperref[fig:#1]{Figure~\ref*{fig:#1}}}
\newcommand{\tab}[1]{\hyperref[tab:#1]{Table~\ref*{tab:#1}}}
\newcommand{\lem}[1]{\hyperref[lem:#1]{Lemma~\ref*{lem:#1}}}
\newcommand{\propos}[1]{\hyperref[propos:#1]{Proposition~\ref*{propos:#1}}}
\newcommand{\thm}[1]{\hyperref[thm:#1]{Theorem~\ref*{thm:#1}}}
\newcommand{\alg}[1]{\hyperref[alg:#1]{Algorithm~\ref*{alg:#1}}}

\newcommand{\IB}[1]{ {\color{magenta} IB: #1}}

\SetCommentSty{mycommfont}

\SetKwInput{KwInput}{Input}                
\SetKwInput{KwOutput}{Output}              

\title{TCBP-JOBS: GPU-Accelerated Quantum Error Correction Decoding and Noise Learning with Near-Optimal Accuracy}

\author{Ben Howe\thanks{Corresponding author: \href{mailto:bhowe@nvidia.com}{\texttt{bhowe@nvidia.com}}.}}
\author{Satvik Maurya\thanks{Corresponding author: \href{mailto:satvikm@nvidia.com}{\texttt{satvikm@nvidia.com}}.}}
\author{Jan Olle}
\author{Ivan Basov}
\author{Muyuan Li}
\author{Justin G.~Lietz}
\author{Kevin Young}
\author{Krysta M.~Svore}
\affil{NVIDIA Corporation}
\date{}

\begin{document}
\maketitle

\begin{abstract}

Achieving near-optimal quantum error correction (QEC) decoding accuracy often requires computationally expensive searches that limit both large-scale simulation and real-time operation. We introduce trajectory-calibrated belief propagation (TCBP), which
learns calibrated fault probabilities from BP trajectories, and jittered OSD beam search (JOBS), which uses
these probabilities to guide parallel GPU searches. Across surface, color and bivariate-bicycle code memory and surgery
circuits, 
TCBP-JOBS achieves logical error rates comparable to those of the Tesseract most-likely-error decoder
at physical error rates of 0.1\% to 0.4\%,
and up to 4.5$\times$ lower on the largest codes. Serial speedups over Tesseract's sparsified presets
range from 10$\times$ to 7,400$\times$, and against unsparsified long-beam Tesseract they reach
13,000$\times$ on the bivariate-bicycle codes, at logical error rate ratios of 0.97 to 1.57.
At lower physical error rates ($10^{-4}$ to $5\times10^{-4}$), rare-event splitting on the gross code shows that Tesseract is more accurate.
Batched execution accelerates offline simulation for rapid, efficient exploration of large-scale QEC design spaces. 
For QPUs with longer logical clock cycles, our serial memory and logical-surgery experiments demonstrate accuracy comparable to Tesseract while meeting the millisecond-scale latency requirements of online feed-forward operation.
Beyond memory and logic decoding, we demonstrate using TCBP-JOBS for confidence scores for postselection in memory and magic-state cultivation,
enabling end-to-end color code cultivation. Together, these methods provide a GPU
framework for accelerating QEC research via accurate decoding, noise estimation, and confidence assessment.
\end{abstract}

\input{inputs/01introduction}
\input{inputs/02background}
\input{inputs/03TCBP}
\input{inputs/04JOBS}
\input{inputs/05noise-learning}
\input{inputs/06complimentary-gap}
\input{inputs/07importance-sampling}
\input{inputs/08conclusions}

\appendix
\section{Supplementary tables}
\input{inputs/A1jobs-vs-tesseract-per-basis}
\FloatBarrier

\input{inputs/A2serial_latency}
\FloatBarrier
\input{inputs/A3batch_throughput}
\FloatBarrier
\input{inputs/A4splitting.tex}
\FloatBarrier

\bibliographystyle{alpha}
\bibliography{references}

\end{document}

%% file: inputs/01introduction.tex
\section{Introduction and Summary of Main Results}
\label{sec:intro}

\sloppy
Quantum computers promise to solve problems that are intractable for classical machines, most notably
the simulation of quantum systems in chemistry and materials science~\cite{lloyd1996universal,
aspuru2005simulated, reiher2017elucidating, lee2021even}, along with a growing set of problems with credible
super-quadratic speedups~\cite{kammerlander2026compendium} and cryptographic tasks such as factoring large
integers~\cite{shor1997polynomial, gidney2025how}, but this promise rests on quantum error correction (QEC).
Physical qubits fail at rates of around $10^{-3}$ per operation, while useful fault-tolerant quantum algorithms are expected to require logical error rates of $10^{-10}$ or lower.
QEC bridges this gap by encoding logical qubits redundantly across many physical qubits and repeatedly measuring parity checks whose outcomes, called syndromes, reveal information about underlying errors while preserving the logical quantum state.
The need for fault-tolerant quantum computing (FTQC) has prompted industry roadmaps that target fault-tolerant machines
within this decade~\cite{google2024roadmap, ibm2024roadmap, quantinuum2025roadmap}, and the QEC codes these
machines will use are still being discovered. While the surface code~\cite{fowler2012surface} has been the
default choice for over a decade, recently quantum low-density parity-check (qLDPC) codes, such as the bivariate-bicycle codes~\cite{bravyi2024high}, have emerged that encode many logical qubits per block at a fraction of the physical-qubit overhead.
These high-rate codes continue to inspire new code constructions, syndrome-extraction circuits and hardware co-designs aimed at reducing the cost of fault tolerance~\cite{panteleev2021degenerate, breuckmann2021qldpc, yoder2025tour, hong2026quantumldpccodesdesign, bhardwaj2026highrateqldpcprocessors}.  To accelerate progress toward practical fault-tolerant quantum computing, rapid exploration of a broader fault-tolerant design space spanning codes, architectures, and hardware co-designs is needed.

At the core of QEC are decoders---classical algorithms that infer likely error corrections from observed syndromes. 
Decoders serve two distinct but complementary roles in quantum error correction.
Offline, a QEC code and its fault-tolerant protocols are evaluated by simulating their circuit implementations under realistic noise and decoding millions to billions of
sampled syndromes. 
These studies are used to estimate logical error rates, characterize performance over many rounds of QEC, assess the impact of different physical noise models, and compare competing circuit implementations and decoder configurations.
Decoders are central to QEC: their accuracy determines the performance achievable by a QEC code, while their computational performance determines the scale and speed of QEC design-space exploration.
Online, once QEC codes are implemented in the quantum processor, the decoder must process syndromes and return corrections within the latency constraints of the target hardware architecture for the computation to proceed~\cite{terhal2015qec}. Decoding is
NP-hard in general~\cite{iyer2015hardness}, so both uses depend on decoders that are simultaneously fast and accurate.

\begin{figure*}
  \centering
  \includegraphics[width=\linewidth]{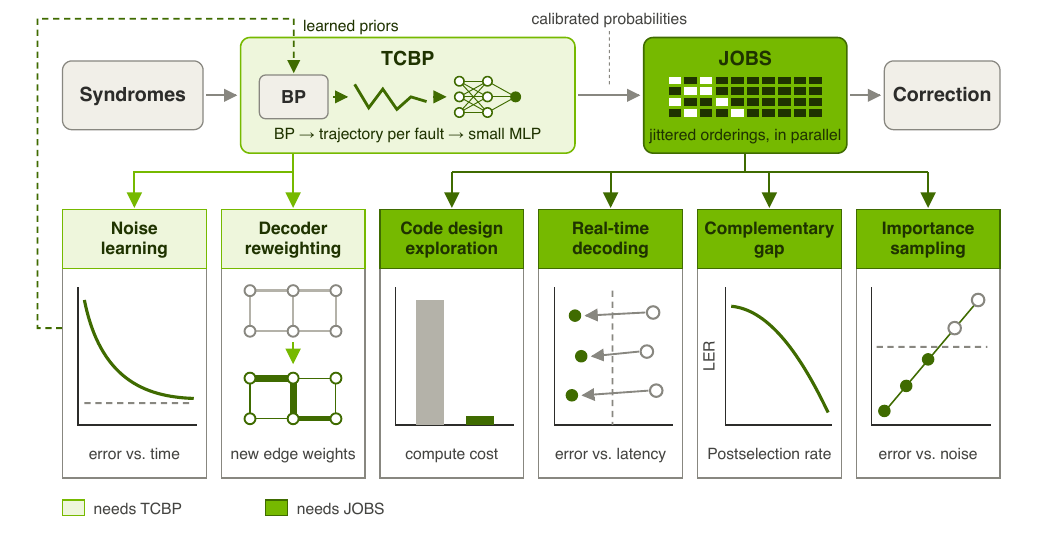}
  \caption{Overview of TCBP-JOBS: A short run of BP produces a trajectory of beliefs for every fault,
  which TCBP maps to calibrated probabilities with one small MLP shared by all faults; JOBS uses these
  probabilities to search many jittered OSD orderings in parallel on the GPU and returns the cheapest
  correction it finds. TCBP alone (light green) supports noise learning, whose learned rates feed back
  into BP as priors, and reweighting of downstream decoders such as matching. TCBP+JOBS (dark green)
  enables code design exploration at a fraction of the compute cost of most-likely-error decoding,
  real-time decoding at lower latency for the same logical error rate, post-selection using the
  complementary gap, and importance sampling to estimate logical error rates far below what direct
  sampling can reach. The insets are schematic.}
  \label{fig:intro-1}
\end{figure*}

Decoders for general codes, however, have been either fast or accurate, but not both. 
While belief propagation (BP) and its descendants~\cite{roffe2020decoding, muller2025relay} are fast and apply to any code, BP frequently fails to converge on quantum codes. 
The post-processor most commonly run on the non-converged shots, called ordered statistics decoding (OSD), commits to a single ordered ranking of likely faults  and only searches around those instances. 
An accuracy gap results and in turn changes how codes are evaluated.
For example, when Tesseract~\cite{beni2025tesseract}
is used to decode the 288-qubit circuit of the $[[144,12,12]]$ gross code, its accuracy matches distance-13 to
distance-15 surface codes encoding 12 logical qubits at SI1000 noise and $p=0.1\%$, resulting in a $14$--$19\times$ savings in physical qubits.  
In contrast, decoding with two faster decoders, BP+OSD and correlated matching, results in accuracy on par with only  the distance-11 surface code, yielding only a $10\times$ physical-qubit savings~\cite{beni2025tesseract}. 
The faster decoders sacrifice accuracy on the very qLDPC codes researchers seek to design and improve due to their high rates.

At the other extreme, most-likely-error decoders such as Tesseract deliver up to two orders of accuracy improvement, but require runtimes of milliseconds per round on small surface codes to
seconds per round on qLDPC codes due to the exponential search space. 
The computational performance of such decoders prevents their use in real-time execution for QPUs with both low (milliseconds) and high (microseconds) logical clock speeds~\cite{nvidia2025helios}.
Further, using these highly accurate decoders render QEC research slow, and reduces the number of codes QEC researchers can study or requires changing to lower-accuracy, fast decoders for QEC code design and exploration.
Decoding 24.2 million shots of one Tesseract benchmark across 28 circuits consumes 130,257 CPU thread-hours \cite{tesseract-pr254}, which is equivalent to around 12 weeks on a 64-core server (see Appendix~\ref{app:batch_throughput}). 
Every task of QEC research that requires accurate decoding consumes this same cost, such as estimating very low logical error
rates, computing confidence values for post-selection, and learning the noise of a device from its syndromes.

This work presents two algorithmic approaches that, when used together, result in accuracy comparable to most-likely-error decoding at physical error rates of 0.1\% to 0.4\%, at the speed of a parallel GPU decoder. The approach builds on the observation that BP computes a belief about every fault
at every iteration, which OSD then discards, and typically keeps only one final value per fault and one ordering of the faults.
We introduce \textbf{Trajectory-Calibrated BP (TCBP)}, which replaces BP's final beliefs with calibrated probabilities computed by a single MLP that reads each fault's entire BP trajectory;
the model is fitted once per code family on simulated shots in about 10\,s and does not require experimental data. 
A second technique, called \textbf{Jittered OSD Beam Search (JOBS)}, uses these probabilities as the starting point for a population of searches. 
It perturbs the ordering to form 16 draws per shot (tunable parameter), solving OSD-0 in each ordering,
and then refines every solution with a bounded beam search, with all of it running on the GPU. 
Across 72 surface, color and bivariate-bicycle code circuits at $p = 0.1\%$ to $0.4\%$, TCBP-JOBS is shown to achieve logical error rates comparable to or lower than those of the Tesseract configurations evaluated. 
Against sparsified short-beam Tesseract, it achieves significantly lower LER in 17 of 36 code/noise settings and significantly higher LER in one, with no statistically significant difference in the remaining settings. 
The largest LER reduction is 4.5×, achieved on the gross code at $p=0.2\%$ SI1000 noise. Against sparsified long-beam Tesseract, it achieves comparable or lower LER at p=0.2\% and 0.4\% SI1000 noise, with a maximum reduction of 2.7×. 
It does so while decoding up to 13,000$\times$ faster across all evaluated circuits and keeping
every shot in a narrow, millisecond-scale latency band. 

Figure~\ref{fig:intro-1} shows an overview of  TCBP-JOBS.
TCBP-JOBS shifts the speed–accuracy frontier for general codes, and provides not only a correction but calibrated fault probabilities and a per-shot confidence. We demonstrate the application of TCBP-JOBS to five settings:
\begin{itemize}
  \item \textbf{Code design exploration.} TCBP-JOBS evaluates QEC codes with accuracy comparable to Tesseract at $p = 0.1\%$ to $0.4\%$ on joint XYZ decoding graphs, at up to $\sim$13{,}000$\times$ lower single-shot latencies (see Appendix~\ref{app:serial_latency}). It decodes the 24.2-million-shot
        Tesseract benchmark in 45 minutes on one GB200 GPU with a geometric-mean logical error rate ratio of
        1.05 in its favor; in contrast Tesseract requires 130,257 CPU thread-hours. The ability to evaluate more shots efficiently opens the door to expanded design exploration of codes, circuits and layouts for a given
        hardware platform, while improved decoder accuracy may enable higher logical-qubit yield and larger computations from the same device.
  \item \textbf{Real-time decoding for QPUs with lower logical clock speeds.} 
        TCBP-JOBS delivers logical error rates comparable to Tesseract at $p = 0.1\%$ to $0.4\%$, at millisecond-scale serial latencies
        on a single GPU, making the decoder practical for, e.g., neutral atom and ion trap quantum processors. For twelve rounds of gross-code memory at $p=0.2\%$, decoding takes 4.08\,ms;
        for gross-code logical-surgery measurement, decoding takes 13.5\,ms per shot, and results in
        27 failures versus 41 and 10.2\,s for multi-pass Tesseract.   
  \item \textbf{Noise learning.} The same TCBP model enables learning the error rate of every
        fault from recorded, unlabeled
        detector data alone: on the gross code, 15 minutes of learning on one GPU
        yields a decoder within 2--3\% of the logical error rate obtained with the true rates, and about
        10\,s of learning comes within 1.4--1.8$\times$ of it, versus 9--27$\times$ for a uniform
        prior, which makes it suitable for online use
        alongside calibration.
  \item \textbf{Complementary-gap based post-selection.} Because JOBS visits many valid corrections, it can
        report the gap between the best competing logical class and its answer at almost no extra cost,
        giving a per-shot confidence up to $\sim 2^k \times$ faster than other decoders which require decoding once for each of the $2^k$ observable classes, where $k$ is the number of observables. Discarding
        1\% of shots improves the LER of the gross code by $\sim 28\times$ at $p=0.2\%$. In magic-state cultivation,
        TCBP-JOBS reduces the binary-Y diagnostic error (wrong-sign decoded logical-$Y$ expectation) by 2.1 to 4.7$\times$ under identical postselection rules with the original cultivation and grafted surface-escape protocol~\cite{gidney2024cultivation},
        and enables end-to-end color code cultivation with up to 2.8$\times$ higher modeled throughput per
        installed qubit.
  \item \textbf{Importance sampling.} Fast decoding makes rare-event splitting affordable: we estimate
        the logical error rate of TCBP-JOBS on the gross code down to $10^{-12}$ at a fraction
        of the cost of direct sampling. In this regime Tesseract is more accurate: the gap is visible at
        $p = 5\times10^{-4}$, $2\times10^{-4}$ and $10^{-4}$, and TCBP-JOBS fails on some four-fault
        configurations that Tesseract corrects.
\end{itemize}

The paper is organized as follows. 
In Section~\ref{sec:tcbp}, we introduce Trajectory-Calibrated BP. We show that BP's ranking of faults is often poor exactly when OSD is needed, and that a small MLP over BP trajectories improves it, reducing OSD-0 failures by 2.1$\times$ on a surface code and 3.7$\times$ on the gross code with only the ordering        changed. In general, TCBP enables the weights of the decoding graphs to be         updated, which can improve the performance of a wide range of decoders. 
In Section~\ref{sec:jobs}, we introduce Jittered OSD Beam Search. 
We present a GPU-parallel search over perturbed OSD orderings        that reaches accuracy at parity with or better than Tesseract at $p = 0.1\%$ to $0.4\%$, with millisecond-scale serial latency.
We use the TCBP model for label-free noise learning        and JOBS for a search-limited complementary gap that enables post-selection in memory experiments and magic-state cultivation in Sections~\ref{sec:noise-learning}        and~\ref{sec:complementary-gap}.
Finally, in Section~\ref{sec:importance-sampling}, we use rare-event splitting to probe logical error rates of $10^{-12}$,
and find that TCBP-JOBS is less accurate than Tesseract at these low physical error rates.

%% file: inputs/02background.tex
\section{Background and Related Work}
\label{sec:background}
This section provides a brief overview of Quantum Error Correction (QEC) and decoding of QEC codes.
We describe the challenge of real-time decoding, and prior art on decoding algorithms.

\subsection{Quantum Error Correction and Decoding}
Qubits are fragile: every gate, idle step, and measurement can introduce an error.
QEC protects information by redundantly encoding $k$ logical qubits into $n > k$ physical qubits.
A code's distance $d$ describes the distance between logical codewords, which quantifies how many errors can be reliably detected.
Stabilizer codes enable stabilizer measurements which remove entropy from the system, and whose outcomes reveal errors.
One goal for QEC codes is to protect the logical qubits by having a high distance, and doing so with the highest rate $k/n$ possible, reducing the physical qubit overhead needed for QEC.
The surface code~\cite{dennis2002topological, fowler2012surface} is a widely studied QEC code which has been demonstrated in experimental settings with excellent error suppression properties.
It requires only nearest-neighbor connectivity but requires a $\sim d^2$ patch of physical qubits per single logical qubit.
High rate Quantum low-density parity-check (QLDPC) codes~\cite{tillich2014hgp, panteleev2021degenerate, breuckmann2021qldpc} encode many logical qubits per block at a fraction of this overhead: the $[[144,12,12]]$ gross code~\cite{bravyi2024high} encodes 12 logical qubits in 144 physical qubits, about $10\times$ fewer than surface codes of the same distance.
This comes with other tradeoffs, such as the need for higher-weight, long-range checks, pushing additional complexity onto hardware realizing these codes.
Despite these challenges, several full architectures have been proposed for enabling large scale FTQC with high rate QLDPC codes ~\cite{yoder2025tour, nvidia2025helios, tripier2026walkingcat, bhardwaj2026highrateqldpcprocessors}.

\textbf{Decoding.} 
Measuring the stabilizers of a QEC code produces syndrome data which is sent to a classical processor called a decoder.
Quantum circuits implementing these stabilizer measurements are called stabilizer rounds, code cycles, syndrome extraction cycles (SECs), or QEC cycles.
They are designed to produce \textit{detectors} which are parities of sets of measurements which are deterministic in the absence of noise ~\cite{gidney2021stim, Derks_2025}.
Decoders operate on a \textit{detector error model}
(DEM)~\cite{gidney2021stim}: a binary matrix $H \in \mathbb{F}_2^{m \times n}$ with a real vector $P \in \mathbb{R}^{n}$ in which each of the $n$ columns is an error mechanism (fault) with prior probability $p_j$ and each of the $m$ rows is a detector.
A second binary matrix $L$ records which logical observables each mechanism flips.
Given the measured syndrome $\sigma$, decoding solves
\[
H F = \sigma
\]
for a fault vector $F$.
Due to quantum degeneracy, the system has many solutions for $F$, and we say the decoder succeeds if $L F$ matches the true observable flips.
Optimal (maximum likelihood) decoding selects the most probable logical class and is \#P-hard~\cite{iyer2015hardness}; most practical decoders instead approximate the \textit{most likely
error} (MLE), the minimum-cost solution with cost $\sum_j w_j F_j$ and $w_j = \ln\frac{1-p_j}{p_j}$,
which is NP-hard in general~\cite{berlekamp1978inherent, iyer2015hardness}.

By viewing each of the rows (checks) and columns (faults) of the DEM as two different node types, we can view the entries of $H$ as an adjacency matrix of a bipartite graph called the Tanner graph.
For the surface code, each mechanism flips at most two detectors (after decomposition), and so the fault nodes of the Tanner graph can be collapsed into edges of a graph of detector nodes.
Here we say $H$ is graph-like and MLE decoding reduces to minimum-weight perfect matching (MWPM)~\cite{edmonds1965paths, dennis2002topological}.
In QLDPC DEMs, mechanisms flip many detectors, and so the adjacency matrix of $H$ describes a hypergraph, and matching does not apply.
The current default general-purpose decoder is BP-OSD~\cite{panteleev2021degenerate,
roffe2020decoding}.
Belief propagation (BP)~\cite{gallager1962ldpc, pearl1988probabilistic} is an iterative algorithm which passes messages on the Tanner graph of $H$ to estimate the posterior probability of every mechanism; if the
hard decision of these posteriors satisfies $\sigma$, BP has \textit{converged}.
On quantum codes, degeneracy and short cycles frequently prevent convergence~\cite{poulin2008iterative}, especially under
circuit-level noise. Ordered statistics decoding (OSD)~\cite{fossorier1995soft} then sorts the columns of
$H$ by BP's posteriors, uses Gaussian elimination to find the first linearly independent set of columns, and solves for the unique correction supported on them (OSD-0).
Higher-order OSD (exhaustive or combination sweep, OSD-CS) flips a parametrized subset of the remaining columns and keeps the most probable valid correction found.
OSD-HO-E(i) (OSD with exhaustive sweep with $i$ additional columns) is a higher-order extension to OSD which takes the OSD-0 solution, and adds all $i$ column subsets to a search for valid (syndrome resolving) fault patterns of higher probability.
By extending $i$ to include all remaining columns, one would recover the (combinatorially expensive) MLE solution.
OSD's answer is only as good as its column ordering: error mechanisms ranked low are unlikely to be recovered even with large sweeps using higher order OSD.

\subsection{Real-time Decoding}
QEC cycles run continuously on the hardware, and every cycle $T_{cycle}$ produces a new round of syndromes,
from every $\sim 1\,\mu s$ on superconducting qubits~\cite{acharya2025below} to every few milliseconds on
trapped ions and neutral atoms~\cite{nvidia2025helios}. A real-time decoder must therefore satisfy three
requirements at once. The first is \textit{accuracy}: a correction succeeds only if it lies in the same
logical class as the error that actually occurred, and every decoding mistake adds to the logical error
rate (LER) of the computation. The second is \textit{throughput}: if syndromes are decoded more slowly than
they are generated, unprocessed syndromes accumulate and the computation suffers an exponential
slowdown---the backlog problem~\cite{terhal2015qec, holmes2020nisq}. The third is \textit{latency}:
non-Clifford gates implemented with magic states require a feed-forward correction that depends on a
decoded logical measurement~\cite{litinski2019game}, so the computation stalls until the decoder
responds~\cite{maurya2026elastic}. Because decoding time depends on the syndrome, the tail of the latency
distribution matters as much as its mean: a single slow decode delays every operation that depends on
it~\cite{delfosse2023choose, maurya2026elastic}.

\subsection{Related Work}
\textbf{Matching and clustering decoders:} MWPM decoders for the surface code have been accelerated in
software~\cite{higgott2023sparse, wu2023fusion} and extended to correlated errors by reweighting
edges~\cite{fowler2013correlated} or with BP-derived weights (belief-matching)~\cite{higgott2023improved}.
Ensembles of matching decoders with perturbed priors approach maximum likelihood accuracy, either by
voting (Harmony)~\cite{shutty2024efficient} or by combining the best local solutions
(Libra)~\cite{jones2024improved}. HyperBlossom~\cite{wu2025minimum} generalizes the blossom algorithm to
hypergraphs with certifiable proximity bounds. Union-find decoders~\cite{delfosse2021almost,
lobl2024breadth, peduzzi2026reducing} grow clusters around syndrome bits until each cluster can be
corrected locally; generalized union-find extends this to qLDPC codes~\cite{delfosse2022toward} using a
GF(2) validity check per cluster. Color codes are decoded by reducing the problem to
matching~\cite{kubica2023efficient, gidney2023new, lee2025color}.

\textbf{Pre-decoders:} Pre-decoders remove simple, local errors before a slower global decoder runs,
reducing the syndrome density it sees~\cite{delfosse2020hierarchical, smith2023local, ravi2023better,
alavisamani2024promatch, caune2023belief}. Local neural networks built from 3D convolutions are well
suited to this role~\cite{chamberland2023techniques}: NVIDIA's Ising pre-decoders are 3D CNNs that run
on GPUs ahead of PyMatching for surface codes~\cite{chamberland2026fast} and Chromobius for color
codes~\cite{olle2026fast}; they reduce both latency and logical error rates, the latter by $347\times$
for distance-31 color codes. QuantiSpect~\cite{gao2026quantispect} factorizes these convolutions to reduce
their cost. These pre-decoders exploit the geometric locality of topological codes; qLDPC codes lack this
locality.

\textbf{Message-passing decoders:} Several variants improve BP's convergence on quantum codes. Memory
BP~\cite{kuo2022exploiting} adds a memory term to the posterior updates, and Relay-BP~\cite{muller2025relay}
chains memory-BP legs with disordered memory strengths, matching or exceeding BP-OSD on bivariate bicycle
(BB) codes without post-processing. GARI~\cite{maan2025decoding} rewrites the correlated DEM to remove
short cycles and runs an ensemble of 24 randomized min-sum decoders, matching Relay-BP. Other approaches
learn the BP update schedule by reinforcement learning~\cite{moradi2026low}, run ensembles of BP decoders
on syndromes transformed by code automorphisms~\cite{koutsioumpas2025autdec}, decimate the least reliable
mechanisms with guessing on a few branches (BP-GDG)~\cite{gong2024gdg}, modify the Tanner graph to break
degenerate symmetries~\cite{ducrest2022stabilizer, yin2024symbreak}, or flip syndrome bits of oscillating
mechanisms and re-run short BP trials in parallel (BP-SF)~\cite{wang2025bpsf, wang2026gpubpsf}. Neural BP
learns message weights from data~\cite{liu2019neural, miao2023neural}.

\textbf{Post-processing for BP:} Besides OSD~\cite{panteleev2021degenerate, roffe2020decoding}, several
post-processors reduce OSD's cost. Some shrink the problem before elimination, by removing reliable
mechanisms~\cite{kung2024efficient}, restricting OSD to high-likelihood columns~\cite{maurya2025fpga}, or
constraining and terminating the sweep early~\cite{liang2025low}. Best-first OSD~\cite{banfi2026best}
traverses the syndrome-consistent solutions in order of increasing cost, matching BP-OSD with
$100\times$ fewer queries. Localized statistics decoding (LSD)~\cite{hillmann2025lsd} performs OSD's
elimination on independent clusters grown from BP's soft information, ambiguity clustering
(AC)~\cite{wolanski2024ac} splits the problem after BP into independently decodable clusters, and the
ordered Tanner forest~\cite{iolius2024otf} and closed-branch~\cite{iolius2024closed} decoders avoid
matrix inversion altogether. With the exception of automorphism ensembles~\cite{koutsioumpas2025autdec},
which require a code with a large automorphism group, these methods commit to a single ordering derived
from one BP run.

\textbf{Search-based and exact decoders:} Integer programming~\cite{landahl2011color} and
MaxSAT~\cite{noormandipour2024maxsat} decoders find the exact MLE, and tensor-network
decoders~\cite{bravyi2014efficient, chubb2021general, lu2026tensormld, cudaqx_tensor_network_decoder} approximate maximum likelihood
decoding; these are generally orders of magnitude too slow for real-time use. Tesseract~\cite{beni2025tesseract,
grbic2026accelerating} finds the MLE with A$^*$ search and pruning heuristics, and shows that MLE decoding
makes the gross code $14$--$19\times$ more efficient than surface codes, compared with $10\times$ under
BP-OSD and correlated matching. Decision-tree decoders~\cite{ott2025decision} search over branches of the
solution space with optimality guarantees. The beam search decoder~\cite{ye2025beam} branches on the least
reliable mechanism, re-runs masked BP on each branch, and keeps a fixed beam of the most reliable paths,
achieving up to $17\times$ lower logical error rates than BP-OSD on the gross code with sub-millisecond
$99.9^{th}$ percentile latency on a single CPU core. Perturbing a ranking and re-decoding is a classical
idea for improving soft-decision decoders, from Chase decoding~\cite{chase1972class} to sampling diverse
candidates with Gumbel perturbations~\cite{kool2019stochastic}.

\textbf{Neural decoders:} Neural networks decode the surface code accurately~\cite{torlai2017neural,
varsamopoulos2017decoding, bausch2024learning}, and AlphaQubit 2~\cite{senior2025scalable} decodes
surface codes up to distance 11 and color codes up to distance 9 in under $1\,\mu s$ per cycle with
near-optimal accuracy. A transformer decoder has been trained on GPUs for magic state distillation on
neutral atoms~\cite{nvidia2025quera}. Recent work extends neural decoders to BB codes with recurrent
transformers~\cite{blue2026machine}, graph neural networks~\cite{hu2025efficient} and code-specific convolutional neural networks~\cite{gu2026scalableneuraldecoderspractical}; these report mixed
results against BP-OSD on the gross code and must be trained for every code and noise model.

\textbf{Hardware decoders:} FPGA, ASIC, and superconducting-logic decoders for the surface code include
LUT-based~\cite{das2022lilliput}, union-find~\cite{das2022afs, liyanage2023scalable},
clustering~\cite{barber2023realtime, ziad2024local}, MWPM~\cite{vittal2023astrea, wu2025micro},
predecoding~\cite{ravi2023better, alavisamani2024promatch}, and cryogenic~\cite{ueno2021qecool} designs, as
well as network-integrated decoding systems~\cite{wu2024lego, liyanage2025network}. For qLDPC codes,
FPGA/ASIC designs exist for BP and OSD~\cite{valls2021syndrome, garciaherrero2025exploring}, Relay-BP
completes a BP iteration in 24\,ns on an FPGA for the gross code~\cite{maurer2025realtime}, GARI decodes
the gross code at 596\,ns per round on a single FPGA~\cite{bascones2026scalable}, learned BP schedules
have been mapped to FPGA and ASIC~\cite{moradi2026hardware}, and Vegapunk~\cite{zhou2025vegapunk}
decouples the check matrix offline to enable sub-microsecond greedy hierarchical decoding. FPGA-tailored
variants of OSD and clustering remain slower and less accurate than Relay-BP under fixed cycle
budgets~\cite{maurya2025fpga}.

\textbf{GPU decoders:} CUDA-Q QEC~\cite{cudaqx} provides GPU-accelerated BP, Relay-BP, and BP-OSD, and
was used for the first real-time decoding of a qLDPC code~\cite{nvidia2025helios}; its batched BP-OSD-0
has also been used to run automorphism ensembles on a GPU~\cite{qeccodes2025autdec}. Markidis et
al.~\cite{markidis2026engineering} run Relay-BP followed by an on-device OSD-0 on data-center and
commodity GPUs, Wang et al.~\cite{wang2026gpubpsf} run BP-SF trials concurrently on the GPU, and
TensorMLD~\cite{lu2026tensormld} contracts compressed tensor networks on GPUs and FPGAs. Neural decoders
and pre-decoders run on GPUs and TPUs by design~\cite{chamberland2026fast, olle2026fast,
senior2025scalable}. To the best of our knowledge, no prior GPU decoder runs higher-order OSD, or a search
beyond OSD-0, entirely on the device.

\textbf{Soft outputs and post-selection:} A decoder can report how close the best competing logical
class is to its answer---the complementary gap---which predicts logical failures and enables
post-selection~\cite{bombin2022fault, gidney2023yoked, smith2024mitigating, meister2024efficient}. Exact
gaps require one constrained decode per competing class, so practical decoders approximate them, for
example from the agreement among an ensemble~\cite{shutty2024efficient}.

%% file: inputs/03TCBP.tex
\section{Trajectory-Calibrated Belief Propagation}
\label{sec:tcbp}

\begin{figure*}[t]
  \centering
  \begin{subfigure}[b]{0.25\textwidth}\centering\includegraphics[width=\linewidth]{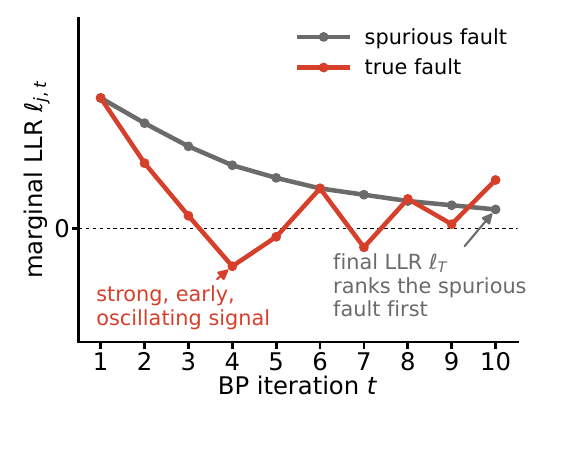}\caption{BP marginal trajectories}\end{subfigure}\hfill
  \begin{subfigure}[b]{0.50\textwidth}\centering\includegraphics[width=\linewidth]{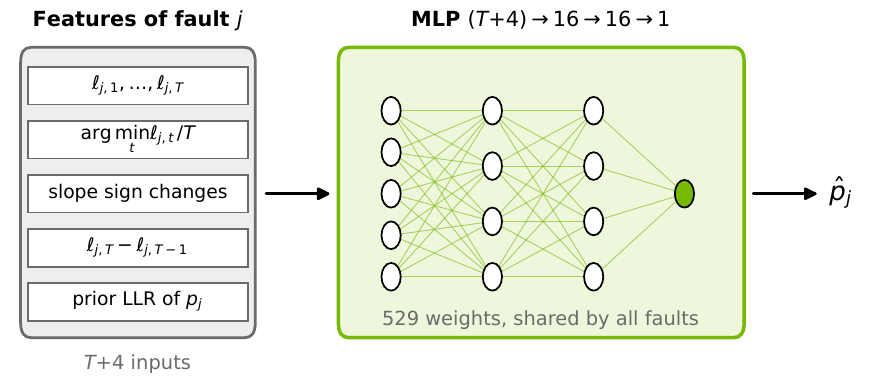}\caption{One shared MLP per fault}\end{subfigure}\hfill
  \begin{subfigure}[b]{0.22\textwidth}\centering\includegraphics[width=\linewidth]{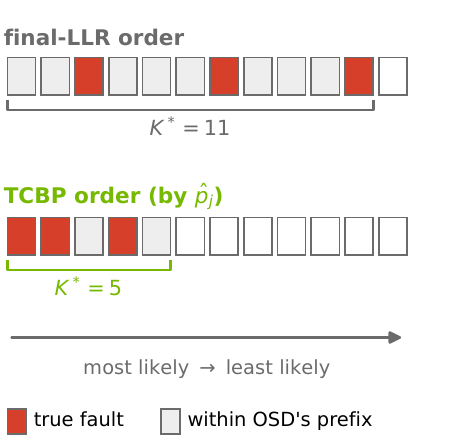}\caption{OSD column ranking}\end{subfigure}
  \caption{Trajectory-Calibrated BP (TCBP). (a) BP updates the marginal LLR of every fault at every
  iteration; a true fault can produce a strong but oscillating signal and finish with a higher LLR than a
  spurious fault that settles slowly (illustrative trajectories). (b) TCBP builds $T{+}4$ features from
  each fault's trajectory and applies one small MLP, shared by every fault, to obtain a calibrated
  probability that the fault fired. (c) Ordering the columns by these probabilities places the true faults
  earlier, so OSD needs a shorter prefix $K^*$ of the ordering to explain the syndrome.}
  \label{fig:tcbp-overview}
\end{figure*}

Decoders like OSD rely on BP to rank the faults, and this ranking is often poor precisely when OSD is needed: OSD only
runs on shots where BP fails to converge, which under circuit-level SI1000 noise is $92.5\%$ of the shots of a
distance-11 surface code at $p=0.2\%$ and every shot of the gross code at $p=0.3\%$
(Section~\ref{sec:tcbp-eval}). On these shots BP's marginals oscillate or stall on trapping sets created by
degeneracy and short cycles~\cite{poulin2008iterative}, so the final LLR of a fault reflects where BP
happened to stop rather than how likely the fault is. Figure~\ref{fig:tcbp-overview}(a) shows a typical
failure: a true fault produces a strong, early dip that BP later abandons, while a spurious fault drifts
slowly towards a small final LLR and is ranked ahead of it. Ordering by the minimum LLR over all iterations
recovers part of this information~\cite{cudaqx}, but it records only the depth of an excursion and not its
timing or stability. Our key observation is that the whole trajectory of a fault's marginal LLRs carries
far more information about whether that fault occurred than any single value taken from it, and that the
mapping from trajectories to probabilities can be learned once for a code family.

\textbf{Model.} For a fault $j$ and $T$ BP iterations, TCBP reads the marginal LLRs
$\ell_{j,1},\ldots,\ell_{j,T}$ and appends four features: the iteration of the minimum LLR normalized by
$T$, the number of sign changes of the successive slopes, the final slope $\ell_{j,T}-\ell_{j,T-1}$, and
the prior LLR of the fault. A multilayer perceptron with two hidden layers of 16 units maps these $T{+}4$
features to a logit, whose logistic transform is the calibrated probability
$\hat{p}_j=P(e_j=1\mid \ell_{j,1:T},p_j)$ (Figure~\ref{fig:tcbp-overview}(b)). The same MLP is applied
independently to every fault, so for $T=10$ the entire model has only 529 weights, its cost is a few
hundred multiply-adds per fault, and it parallelizes trivially across the columns of $H$ on a GPU. BP
itself is unchanged: TCBP only taps the marginals BP already computes, and OSD then sorts the columns by
$\hat{p}_j$ instead of by the final LLR (Figure~\ref{fig:tcbp-overview}(c)).

\textbf{Training.} TCBP needs no experimental data. The model is fitted on shots simulated from $H$ alone,
with a training error rate drawn log-uniformly between $10^{-4}$ and $10^{-2}$ for every fault, so the
labels are known and the model learns how trajectory shapes relate to error probabilities rather than the
rates of one operating point; the BP configuration used for fitting (method, iteration count, clipping and
precision) must match the one used for decoding. Fitting took $9.8$\,s for the surface code DEM with
$24{,}485$ faults and $11.5$\,s for the gross code DEM with $67{,}752$ faults on one NVIDIA H100, and
reduced the held-out cross-entropy of the per-fault probabilities from $0.0053$ to $0.0034$ and from
$0.0107$ to $0.0092$ respectively. A model fitted once can be reused across the error rates and distances
of a code family.

\begin{figure*}[t]
  \centering
  \begin{subfigure}[b]{0.32\textwidth}\includegraphics[width=\linewidth]{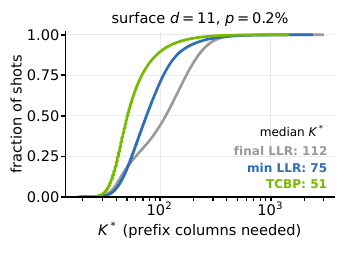}\caption{}\end{subfigure}\hfill
  \begin{subfigure}[b]{0.32\textwidth}\includegraphics[width=\linewidth]{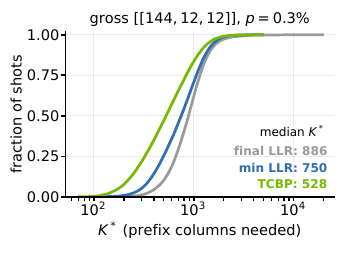}\caption{}\end{subfigure}\hfill
  \begin{subfigure}[b]{0.32\textwidth}\includegraphics[width=\linewidth]{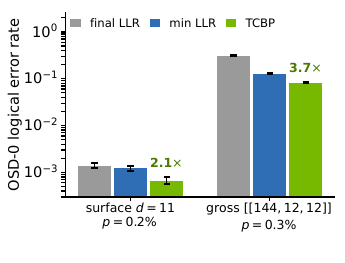}\caption{}\end{subfigure}
  \caption{OSD-0 with three column orderings from the same 10 iterations of BP, on identical
  circuit-sampled shots. (a, b) Distribution of $K^*$, the smallest prefix of the ordering whose columns
  can explain the syndrome, over BP-non-converged shots. (c) OSD-0 logical error rate with 95\% Wilson
  intervals; labels give the reduction of TCBP relative to the final-LLR ordering.}
  \label{fig:tcbp-osd0}
\end{figure*}

\subsection{TCBP Improves the Ranking Given to OSD}
\label{sec:tcbp-eval}
To isolate the effect of the ranking, we decode identical circuit-sampled shots with the same BP front end
(10 iterations of sum-product BP, clip value 15, FP32) and OSD-0, and change only the column ordering
between the final LLR, the minimum LLR and TCBP. We compare for the distance-11 rotated surface code with 11 rounds
of SI1000 noise at $p=0.2\%$ ($204{,}800$ shots) and the $[[144,12,12]]$ gross code with 12 rounds at
$p=0.3\%$ ($32{,}768$ shots)~\cite{beni2025tesseract}. We report two metrics. The first, $K^*$, measures
the ranking itself and uses only what the decoder sees: for each shot and ordering, $K^*$ is the smallest
number of leading columns whose span contains the measured syndrome, i.e., the shortest prefix of the
ordering in which OSD-0 can find a syndrome-compatible correction. The correction OSD-0 returns lies
entirely within these $K^*$ columns, so a smaller $K^*$ means that OSD-0 explains the syndrome with fewer
and higher-ranked faults. $K^*$ does not consult the simulated faults and does not require them to lie
within the prefix; it is therefore only a proxy for accuracy, since a short prefix can still yield a
correction in the wrong logical class. The second metric is the logical error rate, for which we compare the observable flips of the OSD-0 correction with those of the faults the simulator actually sampled. 

Figure~\ref{fig:tcbp-osd0}(a, b) shows that TCBP shifts the entire distribution of $K^*$ to the left, reducing its median from 112 (final LLR) and 75 (minimum LLR) to 51 on the surface
code, and from 886 and 750 to 528 on the gross code. This translates directly into accuracy when measured
against the simulated faults (Figure~\ref{fig:tcbp-osd0}(c)): OSD-0 fails on 136 surface code shots with the TCBP ordering compared
with 285 for the final LLR and 248 for the minimum LLR; it fails on $8.2\%$ of gross code shots compared with
$30.8\%$ and $12.9\%$. 
The improvement is consistent shot by shot---on the gross code, $7{,}832$ shots
fail only with the final-LLR ordering while $427$ fail only with TCBP---and it added only $5$--$8\%$ to the batched OSD-0 decode time for the runs. 
A better ranking does not remove OSD's dependence on a single ordering;
Section~\ref{sec:jobs} uses these calibrated scores as the starting point for a search over many orderings.

\subsection{TCBP Provides Calibrated Outputs}

The previous section described how fault rankings can be improved by TCBP, which is
useful for OSD, but the outputs provide more information than just a ranking: they
can be interpreted as a calibrated probability. That is, if TCBP's output for a given
fault is $0.1$, that can be interpreted to mean that there is a $10\%$ chance that the
fault actually occurred on this shot. This can be used for per-shot downstream decoder fault
reweighting (similar to Belief Matching \cite{higgott2023improved}), or it can be
used for noise learning as described in Section~\ref{sec:noise-learning}.

To test the ability of TCBP to return calibrated output probabilities, we again
use the gross code at $p=0.3\%$, collecting TCBP outputs and binning the outputs
in log space, and then comparing the predicted error probability of each bin with
the fraction of mechanisms in that bin that were actually flipped in the sampled
error. A perfectly calibrated decoder places every bin on the line $y=x$.

\begin{figure}[H]
  \centering
  \includegraphics[width=0.55\linewidth]{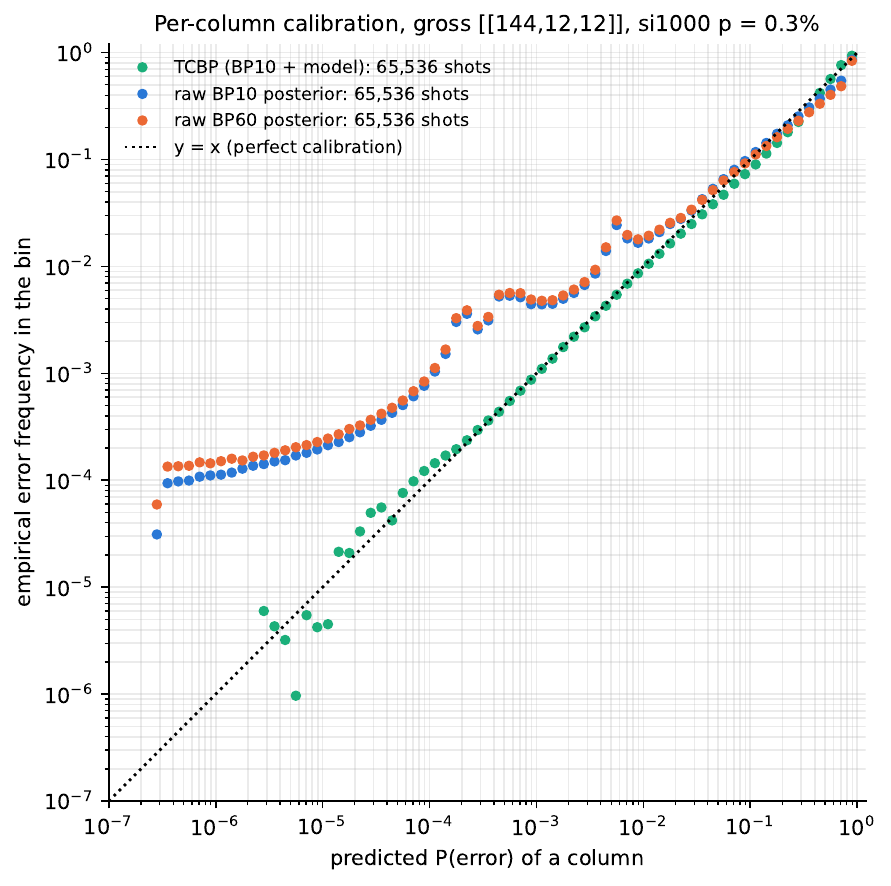}
  \caption{Per-column calibration of TCBP and raw BP posteriors on the gross code at
  $p=0.3\%$. Every mechanism of every shot is one sample ($65{,}536$ shots $\times$
  $67{,}752$ mechanisms per series). Predicted $P(\mathrm{error})$ is binned in log space
  (10 bins per decade) and plotted against the fraction of mechanisms in the bin that were
  flipped in the sampled error; a calibrated decoder places every bin on the dotted $y=x$
  line. Bins with no observed errors are omitted. Raw BP posteriors are
  $1/(1+e^{\mathrm{LLR}})$ from the final iteration with message clipping
  at 15 (lower bound $\approx 3\times 10^{-7}$).}
  \label{fig:tcbp-calibration}
\end{figure}

Figure~\ref{fig:tcbp-calibration} shows that TCBP's outputs track the diagonal
across the range where nearly all faults occur, with only a mild overconfidence at
the high end. Below $10^{-4}$, outside the range of rates the model was trained on,
the outputs drift from the diagonal, but these bins contain well under $1\%$ of the
sampled faults. The raw BP posteriors are calibrated only above $10^{-1}$; below
that they overstate confidence by one to two orders of magnitude, and running more
iterations does not help. TCBP's outputs can therefore be used directly as
probabilities, not just as a ranking. The effects of this capability will be discussed
in Section~\ref{sec:noise-learning}.

%% file: inputs/04JOBS.tex
\section{Jittered OSD Beam Search}
\label{sec:jobs}
\providecommand{\todo}[1]{\IB{TODO: #1}}


Section~\ref{sec:tcbp} improved the ranking that OSD receives, but conventional OSD only explores solutions
that are derived from that one ranking.
OSD~\cite{fossorier1995soft,roffe2020decoding} sorts the columns of $H$ from the most to the
least likely fault, and Gaussian elimination selects the first linearly independent columns
in that order as a basis $B$. The OSD-0 correction solves $H_B e_B = s$ for the syndrome $s$
and sets every other fault to zero. Higher-order OSD, typically the combination
sweep~\cite{roffe2020decoding}, then tries to lower the cost by switching on one or two of
the remaining free columns at a time and solving again for the basis part. It stops there:
the sweep never combines more than two free columns, and every candidate is built on the
same basis. If the true faults fall late in an imperfect ordering, or the cheapest
correction differs from OSD-0 in more than two free columns, the combination sweep does not find it. Jittered OSD Beam Search (JOBS) replaces this single path with a
population of searches, each started from a different plausible ordering, that run in
parallel on a GPU and return the cheapest correction any of them finds.
Algorithm~\ref{alg:jobs} lists the steps for one shot.

\begin{algorithm}[H]
\caption{JOBS for one BP-non-converged shot (one round of draws).}
\label{alg:jobs}
\KwIn{$H$, syndrome $s$, ordering scores $\sigma_j$ (smaller is more likely), priors $p_j$
      with cost $C$ of Eq.~\eqref{eq:jobs-cost}; draws $D$, noise scale $\beta$, window
      pad $\alpha$, beam width $W$, depth $L$, patience $P$}
\KwOut{correction $e^*$ with $He^* = s$}
$K^* \gets$ smallest prefix of the columns sorted by $\sigma$ whose span contains $s$\;
$M \gets \lceil \alpha K^* \rceil$ \tcp*{$M \ge K^*$, so the window of draw 0 spans $s$}
\For{$d = 0, \ldots, D-1$ \textbf{in parallel}}{
  $\sigma' \gets \sigma$ if $d = 0$, else $\sigma + \beta g$ with $g_j \sim \mathrm{Gumbel}(0,1)$ seeded from $s$\;
  $H_d \gets$ the $M$ leading columns of $H$ sorted by $\sigma'$\;
  \lIf{$s \notin \operatorname{span}(H_d)$}{\textbf{continue}}
  eliminate $H_d$ in order: basis $B$, and OSD-0 correction $e_d$ with $H_B (e_d)_B = s$\;
  \ForEach{free (non-pivot) column $j$ of $H_d$}{
    $u_j \gets$ the vector with $(u_j)_j = 1$, basis part solving $H_B (u_j)_B = h_j$, and zeros elsewhere\;
  }
  \tcp{$H u_j = h_j + h_j = 0$, so adding $u_j$ to a correction preserves the syndrome}
  $\mathcal{B} \gets \{e_d\}$\;
  \For{$\ell = 1, \ldots, L$}{
    $\mathcal{N} \gets \{\, e \oplus u_j : e \in \mathcal{B},\ j \text{ free} \,\}$ \tcp*{flip $j$, compensate on $B$}
    $\mathcal{B} \gets$ the $W$ cheapest distinct, not previously visited corrections in $\mathcal{N}$\;
    $e_d \gets \arg\min_{e \in \mathcal{B} \cup \{e_d\}} C(e)$\;
    \lIf{$e_d$ has not improved for $P$ steps}{\textbf{break}}
  }
}
\Return{$e^* = \arg\min_d C(e_d)$ over the draws that were not discarded}\;
\end{algorithm}

\paragraph{Jittered orderings.}

For each shot, JOBS takes one score $\sigma_j$ per fault, here the calibrated TCBP
probability written as an LLR, $\sigma_j = \log\frac{1-\hat{p}_j}{\hat{p}_j}$, so that a
smaller score means a more likely fault, and forms $D$ orderings, or draws.
Draw zero sorts the columns by the score itself, so the search
always contains the ordering that OSD would have used. Each of the remaining $D-1$ draws adds independent
Gumbel noise~\cite{kool2019stochastic}, scaled by $\beta$, to every score and sorts the
columns again, so likely faults still tend to come first.
Two faults with nearly equal scores exchange places in many draws, while a fault that TCBP
separates clearly from the rest keeps its position. The draws therefore concentrate on the
orderings that the scores cannot tell apart, which is where a single ordering is most likely
to be wrong, and the search becomes more productive as the scores become better calibrated.
The noise is seeded from the syndrome, so a shot is searched with the same draws on every
call and the decoder's output is repeatable.

\paragraph{Search window.}
One important speed optimization is that JOBS does not eliminate over all of $H$.
Before any noise is applied, it computes $K^*$, the
smallest prefix of the unperturbed ordering whose span contains the syndrome
(Section~\ref{sec:tcbp-eval}), and fixes a window size of $\lceil \alpha K^* \rceil$ columns
for the shot, with $\alpha = 8$ by default. The number of columns is shared by all draws, but due to the shuffles, the column contents are not: each draw keeps the leading $\lceil \alpha K^* \rceil$ columns of its own jittered
ordering, so the noise changes which columns a draw may use as well as their order. A
jittered draw whose window does not span faults consistent with the syndrome is discarded; the window of draw zero
contains its $K^*$ columns by construction, so every shot returns at least its draw-zero
correction. The
window adapts the work to the shot: an easy syndrome with a small $K^*$ is searched in a
small matrix, and the smaller $K^*$ that TCBP produces (Figure~\ref{fig:tcbp-osd0}) directly
reduces the cost of every draw.

\paragraph{Beam search from each OSD-0 correction.}
Each draw is eliminated in its own order and yields its own basis and OSD-0 correction. From
that correction JOBS runs a beam search over syndrome-preserving moves. A move flips
one free column $j$ of the draw's window, on or off, and adds to the basis
part the solution $x$ of $H_B x = h_j$, where $h_j$ is column $j$ of $H$, so the result still satisfies $He = s$. Each step
extends every correction in the beam by one move and keeps the $W$ cheapest distinct
results that the draw has not visited before, which prevents the beam from cycling between
the same states. Where the combination sweep stops at two free columns, the beam can
move up to $L$ flips away from OSD-0. The search of a draw stops after $L$ steps, or earlier
once $P$ consecutive steps have not improved its best correction. Corrections are compared by their
log-likelihood cost under the prior,
\begin{equation}
  C(e) = \sum_{j\,:\,e_j = 1} \log\frac{1-p_j}{p_j},
  \label{eq:jobs-cost}
\end{equation}
where $p_j$ is the prior probability of fault $j$ from the detector error model. The TCBP
scores decide where JOBS searches, and the prior decides which candidate wins. Every
candidate satisfies the syndrome, so the search never repairs invalid candidates; it compares
different valid explanations of the same detector data. JOBS returns the correction of
least cost over all draws. We use $D = 16$ draws, $\beta = 1$, beam width $W = 32$, depth
$L = 40$ and patience $P = 8$ throughout, with a single round of draws; the implementation
can repeat the $D$ draws with fresh noise for as long as a round lowers the best cost. JOBS runs only on shots where BP does not
converge; it can optionally search the converged shots as well and return the cheaper of
BP's correction and its own.

\paragraph{Relation to other search decoders.}
Other decoders also replace a single BP decision with a bounded search. BP with guided
decimation guessing~\cite{gong2024gdg} runs several rounds of BP, decimates in each round
the fault it expects to be most likely to have flipped, and keeps both values of that fault
open in a small number of rounds. The beam search decoder of~\cite{ye2025beam} branches in every round on
both values of the least reliable fault, runs BP again on each path with the fixed faults
masked out, and prunes the paths to a beam. Both interleave the search with repeated BP runs.
JOBS runs BP once
per shot and spends its search budget after BP, on orderings and on moves between
corrections that already satisfy the syndrome. The draws of a shot are independent of each
other, as are the shots of a batch, so the elimination and beam stages of all
draws run concurrently on the device; Section~\ref{sec:latency} describes the GPU
implementation.

\paragraph{Complementary gap.}
The search evaluates many valid corrections, each with a cost and a logical class. JOBS
can record the lowest cost seen in every class and report the difference between the two
cheapest classes as a confidence score for the shot. Section~\ref{sec:complementary-gap}
defines this score and uses it for postselection.

The rest of this section evaluates TCBP-JOBS as a hard decoder, first for accuracy against
Tesseract on identical shots and then for latency and throughput. The decoder settings are
listed in Table~\ref{tab:jobs-settings}.

\begin{table}[!htbp]
  \centering\small
  \caption{TCBP-JOBS configuration used throughout: the \texttt{tcbp-jobs} preset of the nv-qldpc decoder, plugin defaults otherwise.}
  \label{tab:jobs-settings}
  \begin{tabularx}{\columnwidth}{@{}l>{\raggedright\arraybackslash}X@{}}
    \toprule
    error model & circuit DEM, undecomposed, loops flattened, disjoint errors approximated; DEM rates as priors \\
    BP & sum-product, 10 iterations, message clip $\pm 15$, fp32 \\
    TCBP model & \texttt{init\_tcbp\_model(H, seed=1)}; training rates log-uniform in $[10^{-4},10^{-2}]$, fitted at the BP settings above \\
    column ordering & TCBP scores (draw 0); draws 1--15 re-sorted after Gumbel noise of temperature 1.0 \\
    window & first $\lceil 8K^*\rceil$ columns, $K^*$ the smallest prefix spanning the syndrome \\
    beam search & width 32, depth 40, patience 8; one round \\
    converged shots & BP's answer kept, not passed to JOBS (also searched in Section~\ref{sec:gap-memory} and Appendix~\ref{app:batch_throughput}) \\
    latency runs & batch size 1, one \texttt{decode()} per shot, host wall clock, one GB200 GPU \\
    throughput runs & batch size 4096 (16{,}384 in Appendix~\ref{app:batch_throughput}) \\
    failure & any observable wrong, or syndrome not reproduced \\
    \bottomrule
  \end{tabularx}
\end{table}

%

\subsection{Accuracy against Tesseract on the same shots}
\label{sec:jobs-vs-tesseract}
TCBP-JOBS is a high-accuracy decoder, providing logical error rates comparable to MLE decoders
at the physical error rates tested in this section; Section~\ref{sec:importance-sampling} examines even lower error rate regimes.
We focus the accuracy analysis in this section relative to Tesseract, another high-accuracy decoder.
The comparison uses the shots from 72 circuits under SI1000 noise (joint XYZ decoding),
namely the rotated surface code at distance 7, 9, 11 and 13, the color code at the same
distances, and the bivariate-bicycle codes $[[72,12,6]]$, $[[90,8,10]]$, $[[108,8,10]]$ and
$[[144,12,12]]$, each in $X$ and $Z$ memory at $p = 0.001$, $0.002$ and $0.004$.
Both decoders work on the joint $XYZ$ decoding graph of each circuit throughout: its full,
undecomposed DEM, with the detectors of the $X$ and the $Z$ checks and every error mechanism
in one decoding problem, including the mechanisms that flip detectors of both types. Neither decoder splits it into separate $X$ and $Z$ problems.

\paragraph{Stim circuits:}
\sloppy All circuits evaluated were derived/generated from the Tesseract repository~\cite{beni2025tesseract}\footnote{\url{https://github.com/quantumlib/tesseract-decoder}, commit
\texttt{54569b4}.}.
For the surface and color codes, we get byte-identical circuits to those in the repository up to $d=11$. 
$d=13$ codes were not present in the repository, so they were generated using the same generators. 
For BB codes, we modified the circuits in the repository such that no physical qubit in the circuit has 
more than one operation on it in one time step by inserting \texttt{TICK} instructions. 
These noiseless circuits were then passed through the SI1000 noise model for different physical error rates.

\paragraph{Decoder configurations:} All Tesseract data in this section uses either the short-beam preset or the long-beam preset, both
with error sparsification enabled. As described in the Tesseract decoder Python presets,
we refer to these as
``short beam, sparsified''\footnote{beam 15 with beam climbing, 16 detector orders, sparsification with
$K = 2$ for the surface code and $K = 3$ for the color and bicycle codes, and the circuit's
undecomposed DEM as error model} and ``long beam, sparsified''\footnote{Same as short beam, sparsified except beam 20, 21 detector orders}. The long-beam setting at the same
sparsification is more accurate but slower: it costs 1.3 to
2 times the decode time per shot at $p = 0.001$, and its
median decode time is 1.2 to 2.4 times the short-beam median at $p = 0.002$ and 1.7 to 5.5
times at $p = 0.004$. The short-beam run decodes each stream from shot
0 until its 25th logical error, and a little past it; we call these shots the prefix of the
stream. The longest streams, at $p = 0.001$,
run to $1.3 \times 10^{8}$ shots per circuit. The long beam was run on every shot of the
prefix at $p = 0.002$ and $0.004$; that comparison follows the main results.
TCBP-JOBS decodes the same prefix with
BP (10 iterations, FP32), JOBS with TCBP column ordering, the same DEM rates as prior, and
batches of 4{,}096 shots on one NVIDIA GB200; JOBS runs only on the shots where BP does not
converge, which is the decoder's default.\footnote{Searching the BP-converged shots as well, tested on
the eight smallest $p = 0.001$ circuits, on every bicycle-code row at $p = 0.001$ and on the
$[[144,12,12]]$ code at $p = 0.002$ and $0.004$ ($3.6 \times 10^{8}$ shots in all), changed five
decisions and none of the ratios.} A shot fails if any logical observable is
wrong. The comparison is paired per shot: we count the shots that only Tesseract gets wrong
and the shots that only JOBS gets wrong, test them with an exact two-sided McNemar test, and
report the ratio of the two logical error rates (LER) on the common prefix. The McNemar test
uses only these discordant shots: if the two decoders were equally accurate, a discordant shot
would be equally likely to count against either of them, so the observed split is compared with
a binomial distribution of probability $1/2$. A small test $p$-value, $p_{\mathrm{McN}}$ in the
tables, means that one decoder fails more often than the other on the same shots.
The Tesseract failing shots come from a run on x86 CPUs; a second run on Grace CPUs
reproduced the same failing shots on all 72 circuits. The two runs stop at different points
past the 25th failure, so they share $6.2 \times 10^{8}$ of the $6.6 \times 10^{8}$ prefix shots.

In addition to the results below that compare against runs with sparsification enabled,
we also report serial decoding latency results against
unsparsified long-beam Tesseract in Appendix~\ref{app:serial_latency} and batched throughput
results in Appendix~\ref{app:batch_throughput}. We note that the LER performance of TCBP-JOBS
is largely at parity or better compared to this Tesseract preset as well.

\paragraph{Results: comparison with sparsified short-beam Tesseract.}
Table~\ref{tab:jobs-vs-tesseract} gives the comparison pooled over the $X$ and $Z$ memory
experiments, and Figure~\ref{fig:jobs-vs-tesseract-ratio} plots the pooled LER ratio against
distance; the per-basis rows are in the appendix. Up to distance 9, and on the
$[[72,12,6]]$ code, the two decoders are at parity at every $p$: the LER ratios
(Tesseract LER divided by TCBP-JOBS LER) are 0.9 to
1.2 and no row is statistically significant enough to show non-parity ($p_{\mathrm{McN}} \ge 0.09$). At $p = 0.002$ and $0.004$
JOBS has the lower LER from distance 10 upward in all three families, and the ratio grows with
distance, reaching 3.0 to 4.5 at distance 12 and 13 at $p = 0.002$. There almost every
discordant shot is a Tesseract failure: 40 shots fail only under Tesseract against 1 only under JOBS
on the $[[144,12,12]]$ code, and 35 against 1 on the surface code at distance 13. At
$p = 0.001$ the result depends on the family. The surface code has a ratio of 2.6 at distance
13 (37 against 1) and 1.2 at distance 11 (18 against 5, $p_{\mathrm{McN}} = 0.01$), and the
color code a ratio of 1.9 at distance 13 (85 against 31) while at parity from distance 7 to
11. The bicycle codes are at parity or slightly below, 0.8 to 1.0, and one row is
significant in Tesseract's favor: the $[[108,8,10]]$ code, with an LER ratio of 0.8 and 13
shots failing only under Tesseract against 32 only under JOBS ($p_{\mathrm{McN}} = 7 \times 10^{-3}$).
A possible explanation is that the residual failures there are few-fault
events that Tesseract's search finds and that TCBP-JOBS occasionally misses on these
degenerate codes; Section~\ref{sec:importance-sampling} finds such events on the gross code at lower
physical error rates. Each pooled row has 50 to 170 Tesseract failures, so a single LER ratio has
a 95\% interval of roughly 0.7 to 1.4 times its value even with the pairing.

\paragraph{Results: comparison with sparsified long-beam Tesseract.}
The short-beam setting is Tesseract's fastest preset, not its most accurate one. At
$p = 0.002$ and $0.004$ the long beam was run at the same sparsification on every shot of
the 48 prefixes, $1.7 \times 10^{6}$ shots, so the long-beam columns of
Table~\ref{tab:jobs-vs-tesseract} and the lower row of
Figure~\ref{fig:jobs-vs-tesseract-ratio} are paired with JOBS on the same shots. Summed
over all codes, the long beam fails 563 shots at $p = 0.002$ and 1{,}005 at $p = 0.004$,
against 699 and 1{,}260 for the short beam and 478 and 811 for TCBP-JOBS; 148 shots fail
only under the long beam against 63 only under JOBS at $p = 0.002$
($p_{\mathrm{McN}} = 5 \times 10^{-9}$), and 433 against 239 at $p = 0.004$. The gain is
again on the largest codes. At $p = 0.002$ the LER ratio is 1.9 to 2.2 at distance 12 and 13
(12 shots fail only under Tesseract against 1 only under JOBS on the $[[144,12,12]]$
code); the other rows are at 1.0 to 1.4 and, apart from $[[72,12,6]]$ (1.2,
$p_{\mathrm{McN}} = 0.04$), not significant. At $p = 0.004$ it is 2.7 and 2.5 on the
$[[144,12,12]]$ and $[[108,8,10]]$ codes (51 versus 7 and 55 versus 9), 2.0 and 1.5 on
the surface code at distance 13 and 11, and 1.4 on the color code at distance 13; the
other rows are at 0.9 to 1.2 and not significant. No row at either noise level is
significant in the long beam's favor.

At $p = 0.001$ the prefixes hold $6.6 \times 10^{8}$ shots and the long beam was not run
on all of them, so those cells of Table~\ref{tab:jobs-vs-tesseract} are empty. 
In summary, TCBP-JOBS is at parity with or more accurate than Tesseract's long-beam
setting at $p = 0.002$ and $0.004$, with ratios of 0.9 to 2.7; at $p = 0.001$ we did not
make this comparison.

\begin{table}[tp]
\centering
\caption{TCBP-JOBS against Tesseract on the same shot prefix, both with the circuit's DEM
rates as prior, $X$ and $Z$ memory pooled. Shots is the prefix decoded by both; fails J and
fails T are the failures of TCBP-JOBS and of Tesseract on it. T-only and J-only are the
shots that fail under one decoder only. Ratio is Tesseract LER over JOBS LER.
$p_{\mathrm{McN}}$ is the exact two-sided McNemar test on the discordant shots. Short beam:
Tesseract's short-beam sparsified setting on the whole prefix. Long beam: the long-beam
sparsified setting on the whole prefix, run at $p = 0.002$ and $0.004$. At $p = 0.001$ it was not run
on the whole prefix (dashes): decoding again only the short-beam failures gives a lower
bound on the long-beam failures, and that bound proved loose at $p = 0.002$ and $0.004$.
The $X$ and $Z$ experiments agree within their
intervals; the per-basis table for the short beam is in the appendix. The Tesseract prefix
ends shortly after its 25th short-beam failure, so its short-beam LER on the prefix is
biased upward by a few percent; the paired columns are not.}
\label{tab:jobs-vs-tesseract}
\footnotesize
\setlength{\tabcolsep}{3pt}
\renewcommand{\arraystretch}{0.94}
\begin{tabular}{llrrrrrrrrrr}
\toprule
 & & & & \multicolumn{4}{c}{short-beam, sparsified} & \multicolumn{4}{c}{long-beam, sparsified} \\
\cmidrule(lr){5-8} \cmidrule(lr){9-12}
code & $p$ & shots & fails J & fails T & \shortstack{T-only /\\J-only} & ratio & $p_{\mathrm{McN}}$ & fails T & \shortstack{T-only /\\J-only} & ratio & $p_{\mathrm{McN}}$ \\
\midrule
$[[72,12,6]]$ & 0.001 & 118k & 66 & 59 & 14 / 21 & 0.9 & 0.31 & -- & -- & -- & -- \\
$[[72,12,6]]$ & 0.002 & 9.8k & 60 & 69 & 17 / 8 & 1.2 & 0.11 & 71 & 18 / 7 & 1.2 & 0.04 \\
$[[72,12,6]]$ & 0.004 & 926 & 145 & 159 & 40 / 26 & 1.1 & 0.11 & 156 & 36 / 25 & 1.1 & 0.20 \\
\midrule
$[[90,8,10]]$ & 0.001 & 26.7M & 88 & 85 & 17 / 20 & 1.0 & 0.74 & -- & -- & -- & -- \\
$[[90,8,10]]$ & 0.002 & 179k & 32 & 50 & 23 / 5 & 1.6 & $9 \times 10^{-4}$ & 38 & 12 / 6 & 1.2 & 0.24 \\
$[[90,8,10]]$ & 0.004 & 819 & 51 & 103 & 65 / 13 & 2.0 & $2 \times 10^{-9}$ & 63 & 30 / 18 & 1.2 & 0.11 \\
\midrule
$[[108,8,10]]$ & 0.001 & 110.6M & 98 & 79 & 13 / 32 & 0.8 & $7 \times 10^{-3}$ & -- & -- & -- & -- \\
$[[108,8,10]]$ & 0.002 & 401k & 21 & 51 & 35 / 5 & 2.4 & $1 \times 10^{-6}$ & 30 & 15 / 6 & 1.4 & 0.08 \\
$[[108,8,10]]$ & 0.004 & 820 & 31 & 115 & 91 / 7 & 3.7 & $<\!10^{-10}$ & 77 & 55 / 9 & 2.5 & $4 \times 10^{-9}$ \\
\midrule
$[[144,12,12]]$ & 0.001 & 221.7M & 55 & 53 & 17 / 19 & 1.0 & 0.87 & -- & -- & -- & -- \\
$[[144,12,12]]$ & 0.002 & 356k & 11 & 50 & 40 / 1 & 4.5 & $<\!10^{-10}$ & 22 & 12 / 1 & 2.0 & $3 \times 10^{-3}$ \\
$[[144,12,12]]$ & 0.004 & 452 & 26 & 121 & 97 / 2 & 4.7 & $<\!10^{-10}$ & 70 & 51 / 7 & 2.7 & $2 \times 10^{-9}$ \\
\midrule
color $d{=}7$ & 0.001 & 200k & 58 & 55 & 6 / 9 & 0.9 & 0.61 & -- & -- & -- & -- \\
color $d{=}7$ & 0.002 & 14k & 94 & 104 & 19 / 9 & 1.1 & 0.09 & 103 & 18 / 9 & 1.1 & 0.12 \\
color $d{=}7$ & 0.004 & 1.4k & 134 & 132 & 22 / 24 & 1.0 & 0.88 & 130 & 22 / 26 & 1.0 & 0.67 \\
\midrule
color $d{=}9$ & 0.001 & 1.2M & 54 & 52 & 12 / 14 & 1.0 & 0.85 & -- & -- & -- & -- \\
color $d{=}9$ & 0.002 & 17k & 51 & 53 & 13 / 11 & 1.0 & 0.84 & 52 & 11 / 10 & 1.0 & 1.00 \\
color $d{=}9$ & 0.004 & 819 & 99 & 95 & 43 / 47 & 1.0 & 0.75 & 87 & 31 / 43 & 0.9 & 0.20 \\
\midrule
color $d{=}11$ & 0.001 & 15.5M & 135 & 134 & 48 / 49 & 1.0 & 1.00 & -- & -- & -- & -- \\
color $d{=}11$ & 0.002 & 25k & 36 & 52 & 28 / 12 & 1.4 & 0.02 & 39 & 11 / 8 & 1.1 & 0.65 \\
color $d{=}11$ & 0.004 & 604 & 74 & 111 & 82 / 45 & 1.5 & $1 \times 10^{-3}$ & 84 & 44 / 34 & 1.1 & 0.31 \\
\midrule
color $d{=}13$ & 0.001 & 30.8M & 60 & 114 & 85 / 31 & 1.9 & $5 \times 10^{-7}$ & -- & -- & -- & -- \\
color $d{=}13$ & 0.002 & 20k & 12 & 53 & 48 / 7 & 4.4 & $1 \times 10^{-8}$ & 26 & 19 / 5 & 2.2 & $7 \times 10^{-3}$ \\
color $d{=}13$ & 0.004 & 451 & 88 & 172 & 131 / 47 & 2.0 & $2 \times 10^{-10}$ & 123 & 85 / 50 & 1.4 & $3 \times 10^{-3}$ \\
\midrule
surface $d{=}7$ & 0.001 & 575k & 46 & 50 & 4 / 0 & 1.1 & 0.12 & -- & -- & -- & -- \\
surface $d{=}7$ & 0.002 & 30k & 54 & 58 & 4 / 0 & 1.1 & 0.12 & 58 & 4 / 0 & 1.1 & 0.12 \\
surface $d{=}7$ & 0.004 & 2.0k & 53 & 58 & 9 / 4 & 1.1 & 0.27 & 58 & 9 / 4 & 1.1 & 0.27 \\
\midrule
surface $d{=}9$ & 0.001 & 13.2M & 140 & 141 & 4 / 3 & 1.0 & 1.00 & -- & -- & -- & -- \\
surface $d{=}9$ & 0.002 & 116k & 51 & 55 & 5 / 1 & 1.1 & 0.22 & 53 & 4 / 2 & 1.0 & 0.69 \\
surface $d{=}9$ & 0.004 & 2.0k & 50 & 57 & 15 / 8 & 1.1 & 0.21 & 55 & 11 / 6 & 1.1 & 0.33 \\
\midrule
surface $d{=}11$ & 0.001 & 59.2M & 69 & 82 & 18 / 5 & 1.2 & 0.01 & -- & -- & -- & -- \\
surface $d{=}11$ & 0.002 & 268k & 39 & 53 & 20 / 6 & 1.4 & $9 \times 10^{-3}$ & 39 & 7 / 7 & 1.0 & 1.00 \\
surface $d{=}11$ & 0.004 & 2.0k & 34 & 63 & 38 / 9 & 1.9 & $2 \times 10^{-5}$ & 51 & 27 / 10 & 1.5 & $8 \times 10^{-3}$ \\
\midrule
surface $d{=}13$ & 0.001 & 182.5M & 23 & 59 & 37 / 1 & 2.6 & $3 \times 10^{-10}$ & -- & -- & -- & -- \\
surface $d{=}13$ & 0.002 & 258k & 17 & 51 & 35 / 1 & 3.0 & $1 \times 10^{-9}$ & 32 & 17 / 2 & 1.9 & $7 \times 10^{-4}$ \\
surface $d{=}13$ & 0.004 & 1.8k & 26 & 74 & 58 / 10 & 2.8 & $2 \times 10^{-9}$ & 51 & 32 / 7 & 2.0 & $7 \times 10^{-5}$ \\
\bottomrule
\end{tabular}
\end{table}

\begin{figure}[t]
\centering
\includegraphics[width=\linewidth]{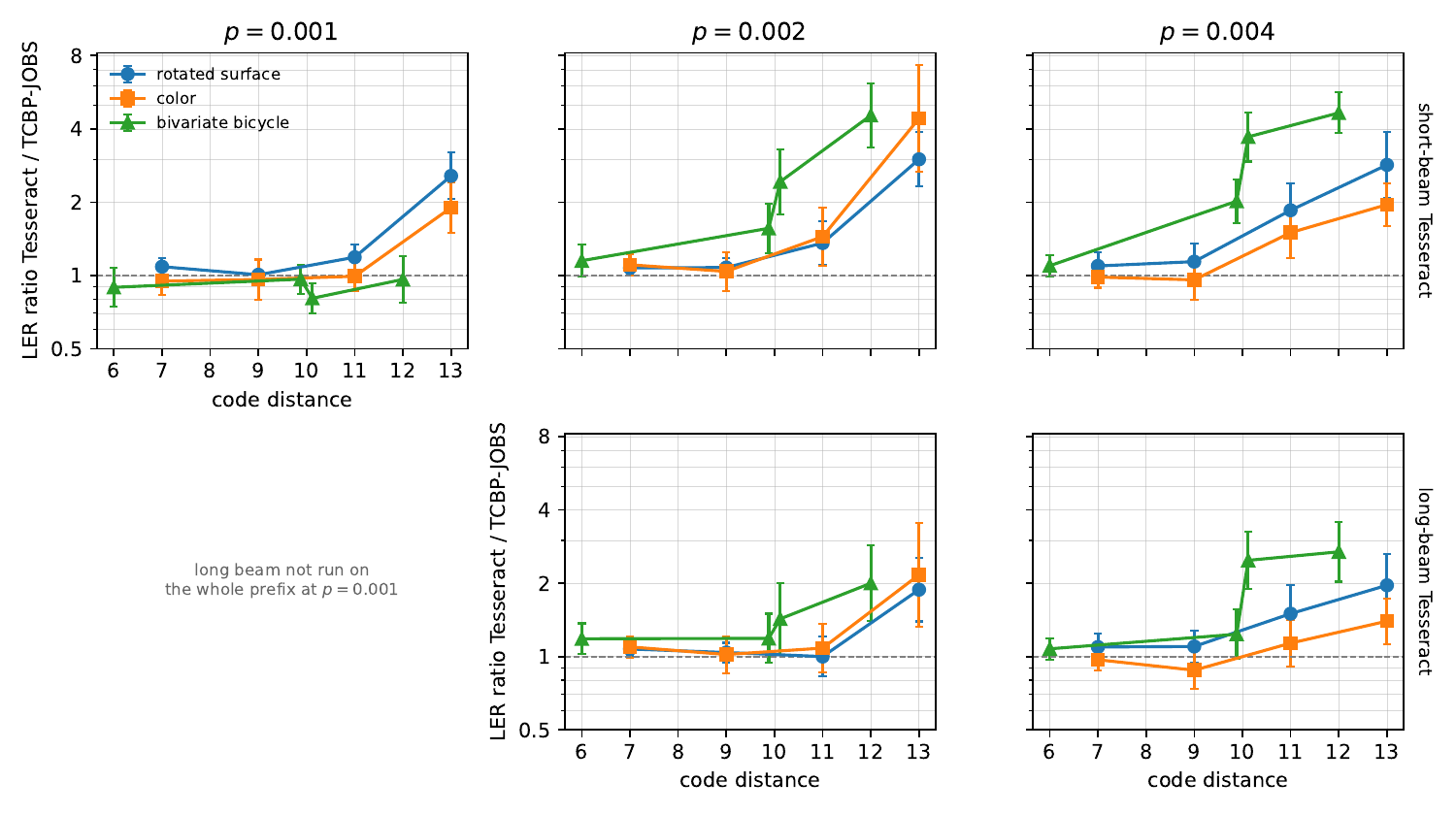}
\caption{Ratio of Tesseract to TCBP-JOBS logical error rate on the common shot prefix
against code distance, one panel per $p$ and one line per code family, $X$ and $Z$ memory
pooled; the two distance-10 bicycle codes are drawn side by side ($[[90,8,10]]$ left,
$[[108,8,10]]$ right). Top row: short-beam sparsified Tesseract; error bars are approximate 95\%
intervals from the paired counts, with the failures common to both decoders held fixed.
Bottom row: long-beam sparsified Tesseract on the whole prefix at $p = 0.002$ and $0.004$, points and
bars as in the top row; the long-beam sparsified was not run on the whole prefix at $p = 0.001$, so
that panel is left out. The dashed line is parity.}
\label{fig:jobs-vs-tesseract-ratio}
\end{figure}

\begin{figure}[t]
\centering
\includegraphics[width=\linewidth]{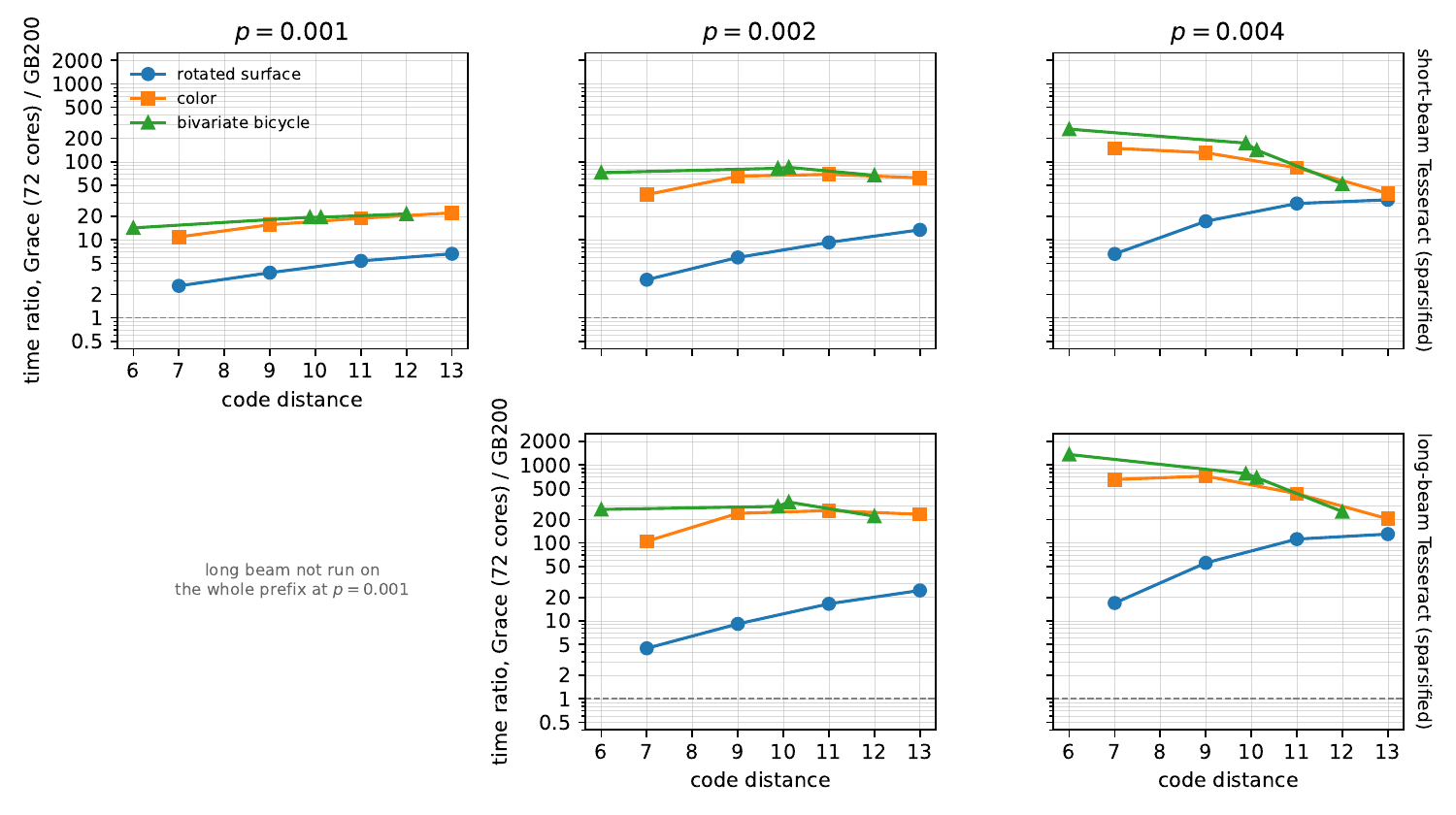}
\caption{Batched throughput ratio of TCBP-JOBS on one GB200 GPU to sparsified Tesseract running on a 72-core Grace CPU
(assuming perfect workload sharing across the CPUs). One panel per $p$ and one line per code family, $X$ and $Z$
memory pooled; the two distance-10 bicycle codes are drawn side by side ($[[90,8,10]]$ left, $[[108,8,10]]$
right). Both are throughput measurements with every core or the whole GPU busy, not latencies
(Section~\ref{sec:latency}).}
\label{fig:jobs-vs-tesseract-throughput}
\end{figure}

\paragraph{Batched cost.}
Figure~\ref{fig:jobs-vs-tesseract-throughput} puts the two decoders on one device each: a 72-core Grace
CPU running sparsified Tesseract with every core busy against one GB200 GPU running TCBP-JOBS in batches
of 4{,}096 shots, both measured as throughput rather than latency. The GPU is 2.6 to 1{,}400 times
faster, the least on the surface codes at $p = 0.001$ and the most on the color and bicycle codes
against the long beam. TCBP-JOBS thus matches or exceeds the accuracy of sparsified Tesseract
(Fig.~\ref{fig:jobs-vs-tesseract-ratio}) at a fraction of the device time. Two other comparisons are more
favorable still: single-shot latency, where the GPU is not amortized over a batch
(Section~\ref{sec:latency}), and Tesseract without sparsification
(Appendix~\ref{app:batch_throughput}), which is sometimes needed to close part of the LER gap to TCBP-JOBS.

\FloatBarrier

\subsection{Potential for real-time decoding}
\label{sec:latency}

A minimum decoding throughput is needed to enable real-time decoding and avoid the
backlog problem~\cite{terhal2015qec,holmes2020nisq}. Parallelizing decoders across
windows can help meet these throughput requirements with additional hardware
resources, subject to the costs of overlapping windows, boundary handling and
communication~\cite{skoric2023parallel,tan2023scalable,bombin2023modular}.
Online decoding also contributes to the reaction time of the classical control
and decoding system. Following the definition in the Supplementary Information,
Sec.~III.D.1 of Ref.~\cite{acharya2025below}, reaction time is the time required
for the control system to respond to a logical measurement by executing a dependent
logical operation. It comprises
\begin{equation}
  T_{\mathrm{react}} = T_{\mathrm{decode}} + T_{\mathrm{control}},
  \label{eq:reaction-time}
\end{equation}
where $T_{\mathrm{decode}}$ is decoder latency and $T_{\mathrm{control}}$ accounts
for gate execution, physical readout and the control system's response to the
decoded outcome.
This reaction time limits the rate of logical operations requiring
feed-forward~\cite{litinski2019game,maurya2026elastic}. The measurements below
characterize decoder-call latency as a contribution to the complete reaction time.

The allowable reaction time depends on the architecture and workload. A
$10\,\mu$s reaction time is assumed in superconducting resource estimates discussed
in Ref.~\cite{caune2024demonstrating}; Google's distance-5 surface-code experiment
demonstrated a mean software decoding latency of $63\,\mu$s~\cite{acharya2025below}.
GPU decoding has also been demonstrated on atomic platforms: BP-OSD on
Quantinuum Helios achieved a $67\,\mu$s median for Bring's code~\cite{nvidia2025helios},
and the NVIDIA--QuEra neutral-atom study reported sub-millisecond decoding for a
distance-3 magic-state-distillation circuit~\cite{nvidia2025quera}. These decoder
measurements establish precedents on their respective workloads; a full-system
reaction-time budget must also account for control and communication.
Here we evaluate whether TCBP-JOBS can deliver accuracy comparable to Tesseract
at decoding latencies relevant to AMO feed-forward operations.

\begin{figure}[!htbp]
  \centering
  \captionsetup{font=small}
  \includegraphics[width=\textwidth]{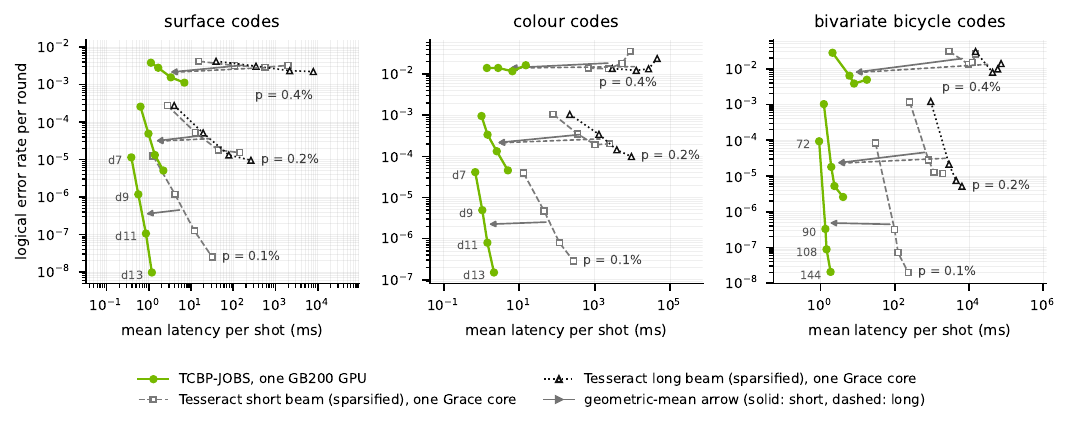}
  \caption{Mean serial decode latency per shot ($d$ rounds) against logical error
  rate per round for the three code families. Each curve joins four code sizes
  at one physical error rate. TCBP-JOBS uses one GB200 GPU and batch size one;
  Tesseract uses one Grace core per process under full CPU load, and both of its
  settings use error sparsification. Accuracy comes
  from paired decoding of common prefixes. Long-beam results are available at
  $p=0.2\%$ and $0.4\%$ only.}
  \label{fig:pareto-family}

  \medskip
  \captionof{table}{TCBP-JOBS serial speedup and Tesseract-to-JOBS LER ratio for
  the memory experiments above. Left: geometric means over four code sizes.
  Right: the code with the largest speedup and its own LER ratio. Both Tesseract
  settings use error sparsification. CPU latencies are measured under full load;
  speedups against an otherwise idle system would be approximately 10--25\% smaller.}
  \label{tab:pareto-family}
  {\small\input{figs/pareto_family_table}}
\end{figure}

Figure~\ref{fig:pareto-family}
compares serial decoder latency with logical error rate across the three code families.
TCBP-JOBS retains the accuracy established in Section~\ref{sec:jobs-vs-tesseract} while
substantially reducing mean serial decoding time. At $p = 0.002$, its geometric-mean
speedups over sparsified short-beam Tesseract are $18\times$, $251\times$, and $376\times$ for surface,
color, and bicycle codes, respectively. Against sparsified long-beam Tesseract, the corresponding
speedups are $30\times$, $866\times$, and $1{,}357\times$.
More importantly, the maximum average decode time across all
memory experiments of $d$ rounds was $18$ ms/shot, even at the near-threshold
error rate of $p=0.4\%$. The mean per-round times ranged from $55\,\mu$s to $1.5$\,ms,
which should be suitable for real-time operation of many AMO systems.
Table~\ref{tab:pareto-family} reports the speedups together with the accompanying
logical-error-rate ratios. 

\paragraph{GPU execution.}
JOBS exploits parallelism within a single shot: its independent perturbed orderings
undergo elimination and beam search concurrently on the GPU. This parallelism remains
available at batch size one. Search effort also adapts to the syndrome. BP-converged
shots retain their correction without invoking JOBS, the search window scales with
$K^*$, and beam patience terminates draws that cease improving. These mechanisms reduce
work on easier shots while allowing harder syndromes more search effort.

\paragraph{Hierarchical decoding for AMO systems.}
Zhao et al.~\cite{zhao2026ultrahighrate} evaluate ultra-high-rate qLDPC memories
for reconfigurable atom arrays using a hierarchy of BP, Relay-BP and an
integer-programming MLE decoder implemented in Gurobi.
Non-converging shots advance to the next tier, making the expensive final stage
rare enough to sustain high simulation throughput.
Their Appendix~C discusses extending this approach to real-time decoding and
identifies faster approximate decoders as candidates to replace exact integer
programming.
A rare slow fallback can nevertheless delay a logical operation that depends on
its result, even when average throughput meets the syndrome arrival rate.
TCBP-JOBS addresses this bottleneck by retaining BP's fast path and applying
TCBP-guided parallel GPU search to the remaining shots.
The measured accuracy--latency tradeoffs here motivate its use as a high-accuracy
fallback in such a hierarchy, potentially replacing or supplementing the
Relay-BP and integer-programming stages.
Our measurements cover different codes and noise models from those of
Ref.~\cite{zhao2026ultrahighrate}; evaluating the replacement on their code family
remains future work, and JOBS does not certify MLE optimality. The complementary gap
from JOBS (Section~\ref{sec:complementary-gap}) could aid in deciding when to pass data
to additional decoding stages.

\paragraph{Timing methodology.}
We measure TCBP-JOBS decode times on one NVIDIA GB200 using CUDA-Q QEC 0.9.0~\cite{cudaqx}, batch size
one, and one \texttt{decode()} call per shot. The host timer surrounds the call and
conversion of the returned result to a NumPy array. Model fitting, decoder construction,
and 200 warm-up calls are excluded. Each circuit contributes the first 20{,}000 shots of
its seeded stream, or the entire accuracy prefix when shorter. Tesseract runs as one
single-threaded process pinned to each physical Grace CPU core, with all cores occupied;
its timing uses the decoder's internal timer. Its distributions cover the first
10{,}000 shots per stream at $p = 0.001, 0.002$, and 1{,}000 at $p = 0.004$.
Accuracy is evaluated separately on the common prefixes described above.

\paragraph{Measured decoder latencies.}
Averaged over the two memory bases, mean TCBP-JOBS latency at $p = 0.002$ ranges from
0.64\,ms for the distance-7 surface code to 4.97\,ms for the distance-13 color code.
The gross code takes 4.08\,ms. At $p = 0.004$, these times increase, reaching 14.9\,ms
for the distance-13 color code and 17.8\,ms for the gross code. Each timing represents
a complete memory shot containing $d$ syndrome-extraction rounds.

\FloatBarrier
\FloatBarrier

\begin{figure}[!htbp]
  \centering
  \captionsetup{font=small}
  \includegraphics[width=\textwidth]{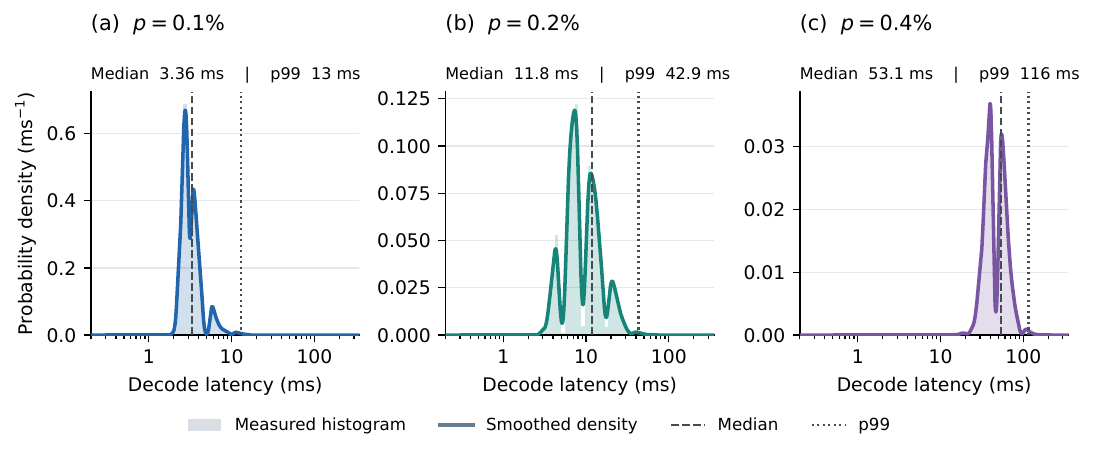}
  \caption{\textbf{TCBP-JOBS serial latency for gross-code $X_1X_2$ logical surgery.}
  Each panel contains 100{,}000 shots, including failures. Shading shows measured
  histograms with logarithmic bins. Curves are Gaussian kernel-density estimates
  in log latency, transformed to density per millisecond; smoothing is for display
  only. Lines mark the empirical median and p99. The horizontal axis is logarithmic
  and vertical scales differ. Panel (c) is a stress test with $55.9\%$ failures.}
  \label{fig:surgery-jobs-latency}

  \medskip
  \captionof{table}{\textbf{Accuracy and serial latency on the surgery shots above.}
  Each row uses the same 100{,}000-shot prefix at its noise rate. Failure includes
  any wrong observable, invalid correction, or Relay-BP non-convergence.
  GPU decoders use one GB200, batch size one; Tesseract uses one Grace core per
  process under full CPU load. Multi-pass Tesseract uses two passes, a long beam
  and a basis-tagged DEM. Timing boundaries are described in the text.}
  \label{tab:surgery-latency}
  {\small\input{figs/surgery_latency_table}}
\end{figure}
\FloatBarrier

\paragraph{Logical-surgery decoder latency.}
Memory experiments characterize decoding accuracy and cost, while logical operations expose the
feed-forward dependencies that make reaction time important. In logical surgery, the decoded outcome
of a joint logical measurement can determine subsequent operations, placing decoding on the
computation's critical path~\cite{litinski2019game}. We therefore evaluate TCBP-JOBS on a simulated intra-block
logical measurement of $\bar{X}_1\bar{X}_2$ on two logical qubits of a single $[[144,12,12]]$ gross
code block, performed by lattice surgery with a graph gadget: 24 vertex checks attached to the support
of $\bar{X}_1\bar{X}_2$ and 50 edge qubits merge the two logicals into one stabilizer, giving a
$[[194,11,12]]$ merged code that is measured for 12 rounds, followed by one round of the original code
and a transversal readout. The circuit uses 383 physical qubits (144 data, 50 edge and 189 ancilla
qubits), 2,481 detectors and 13 observables: the surgery outcome and the 12 logical $\bar{X}$
observables, and a shot fails if any of them is decoded incorrectly. SI1000 noise is applied to steps
2--5; the initial preparation, which includes an ideal reference measurement of the $Z$ checks, and
the final transversal readout are noiseless, following the circuit construction. We decode each shot
over the full circuit and report the software-side decoder latency that contributes to the reaction
time of AMO feed-forward operations.
Table~\ref{tab:surgery-latency} compares four decoders on the same first 100{,}000
shots at each physical error rate. A failure is any wrong observable or an invalid
correction; Relay-BP non-convergence also counts as failure. The stronger Tesseract
baseline uses two passes and a long beam with a basis-tagged DEM.

At $p=0.002$, TCBP-JOBS fails 27 shots against 41 for multi-pass Tesseract, with
mean decoding times of 13.5\,ms and 10.2\,s, respectively. Relay-BP has a lower
median latency than JOBS (6.50\,ms against 11.8\,ms), but fails 2{,}753 shots and
has a longer tail: its p99 is 239\,ms against 42.9\,ms for JOBS. Only $0.375\%$
of JOBS calls exceed 50\,ms, compared with $11.0\%$ for Relay-BP. At $p=0.001$,
neither JOBS nor multi-pass Tesseract fails in these 100{,}000 shots. A longer
paired accuracy run on the same stream finds 10 failures for each on their common
6.3-million-shot prefix, including three failures unique to each decoder.
The $p=0.004$ results are a stress test: both decoders fail on roughly half the
shots, despite JOBS retaining a large latency advantage.

Figure~\ref{fig:surgery-jobs-latency} shows the measured JOBS latency distributions.
The surgery measurements use a batch-one decoder on one GB200 GPU using CUDA-Q QEC 0.9.0~\cite{cudaqx}, with one process
per GPU pinned to four Grace cores, and 20 warm-up calls per 2{,}500-shot piece.
The host timer surrounds only \texttt{decode()}, with a Python-list syndrome
prepared beforehand and NumPy conversion of the result afterward. Tesseract uses
its internal decode timer, with one process per physical Grace core and all cores
occupied. These surgery timings were measured separately from the memory results
above.


\FloatBarrier

%% file: figs/pareto_family_table.tex
\begin{tabular}{llrrrlr}
\toprule
 & & \multicolumn{2}{c}{geometric mean over sizes} & \multicolumn{3}{c}{largest speedup} \\
\cmidrule(lr){3-4}\cmidrule(lr){5-7}
family & $p$ & speedup & LER ratio & speedup & code & LER ratio \\
\midrule
\multicolumn{7}{l}{\emph{Tesseract short beam (sparsified) / TCBP-JOBS}} \\[2pt]
surface & 0.1\% & 10$\times$ & 1.4$\times$ & 27$\times$ & d13 & 2.6$\times$ \\
surface & 0.2\% & 18$\times$ & 1.5$\times$ & 65$\times$ & d13 & 3.0$\times$ \\
surface & 0.4\% & 81$\times$ & 1.6$\times$ & 283$\times$ & d13 & 2.9$\times$ \\
color & 0.1\% & 54$\times$ & 1.1$\times$ & 129$\times$ & d13 & 1.9$\times$ \\
color & 0.2\% & 251$\times$ & 1.6$\times$ & 506$\times$ & d13 & 4.4$\times$ \\
color & 0.4\% & 660$\times$ & 1.3$\times$ & 833$\times$ & d9 & 1.0$\times$ \\
BB & 0.1\% & 71$\times$ & 0.9$\times$ & 123$\times$ & $[[144,12,12]]$ & 1.0$\times$ \\
BB & 0.2\% & 376$\times$ & 2.1$\times$ & 486$\times$ & $[[144,12,12]]$ & 4.5$\times$ \\
BB & 0.4\% & 1,300$\times$ & 2.6$\times$ & 1,607$\times$ & $[[90,8,10]]$ & 2.1$\times$ \\
\midrule
\multicolumn{7}{l}{\emph{Tesseract long beam (sparsified) / TCBP-JOBS}} \\[2pt]
surface & 0.2\% & 30$\times$ & 1.2$\times$ & 118$\times$ & d13 & 1.9$\times$ \\
surface & 0.4\% & 270$\times$ & 1.4$\times$ & 1,126$\times$ & d13 & 2.0$\times$ \\
color & 0.2\% & 866$\times$ & 1.3$\times$ & 1,885$\times$ & d13 & 2.2$\times$ \\
color & 0.4\% & 3,312$\times$ & 1.1$\times$ & 4,580$\times$ & d9 & 0.9$\times$ \\
BB & 0.2\% & 1,357$\times$ & 1.4$\times$ & 1,899$\times$ & $[[108,8,10]]$ & 1.4$\times$ \\
BB & 0.4\% & 6,275$\times$ & 1.8$\times$ & 7,384$\times$ & $[[108,8,10]]$ & 2.6$\times$ \\
\bottomrule
\end{tabular}

%% file: figs/surgery_latency_table.tex
\begin{tabular}{llrrrr}
\toprule
$p$ & decoder & failures & mean (ms) & p50 (ms) & p99 (ms) \\
\midrule
0.1\% & \textbf{TCBP-JOBS} & 0 & 3.95 & 3.36 & 13 \\
0.1\% & Tesseract multi-pass + long & 0 & 420 & 172 & 6,522 \\
0.1\% & Tesseract short + sparsify & 4 & 1,513 & 830 & 12,115 \\
0.1\% & Relay-BP (relay300) & 18 & 1.84 & 1.13 & 10.4 \\
\midrule
0.2\% & \textbf{TCBP-JOBS} & 27 & 13.5 & 11.8 & 42.9 \\
0.2\% & Tesseract multi-pass + long & 41 & 10,237 & 5,322 & 50,458 \\
0.2\% & Tesseract short + sparsify & 2,342 & 20,246 & 19,618 & 46,272 \\
0.2\% & Relay-BP (relay300) & 2,753 & 23.3 & 6.5 & 239 \\
\midrule
0.4\% & \textbf{TCBP-JOBS} & 55,914 & 52.9 & 53.1 & 116 \\
0.4\% & Tesseract multi-pass + long & 47,772 & 102,882 & 103,544 & 144,392 \\
0.4\% & Tesseract short + sparsify & 96,705 & 63,793 & 64,592 & 85,584 \\
0.4\% & Relay-BP (relay300) & 99,867 & 238 & 238 & 244 \\
\bottomrule
\end{tabular}

%% file: inputs/05noise-learning.tex
\section{Noise Learning}
\label{sec:noise-learning}
\providecommand{\todo}[1]{\IB{TODO: #1}}

A decoder needs the probability of every error mechanism in the detector error model (DEM).
On hardware these rates are not known. They drift between calibrations, and a decoder run
with stale or uniform rates loses accuracy. This section shows that the TCBP posteriors
support an unsupervised estimate of all per-mechanism rates from unlabeled detector data, in
minutes on one GPU, and that decoding with the learned rates matches the accuracy of
decoding with the true rates to within a few percent at $p = 0.002$ and $0.004$.

\subsection{Learning procedure}

Estimating error rates from the output of running error detection circuits, instead of
from dedicated characterization experiments such as randomized benchmarking, goes back to
Fowler et al.~\cite{fowler2014scalable}. Our procedure
follows the estimator of Wagner et al.~\cite{wagner2021optimal}, who combine belief
propagation with expectation-maximization: the E-step sums the posterior marginal of every
error over the recorded syndromes, and the M-step sets each rate to that sum divided by the
number of shots. Unlike earlier methods that count the errors in a decoder's hard
decisions, this uses the full posterior of each error. On concatenated codes the factor
graph is a tree, BP marginals are exact, and the estimator is optimal. The authors note
that the method could be applied to qLDPC codes, where the difficulty is that BP is no
longer exact. We take up that case with two changes. First, the E-step uses the calibrated
TCBP posteriors of Section~\ref{sec:tcbp} in place of the BP marginals. Second, the M-step
multiplies each rate by the ratio of its posterior mass on the device data to its posterior
mass on a reference stream simulated from the current rates, which cancels the bias that
remains in the posteriors. With exact posteriors the reference mass equals the expected
number of occurrences of the error and the update reduces to theirs.

\begin{figure}[t]
\centering
\includegraphics[width=\linewidth]{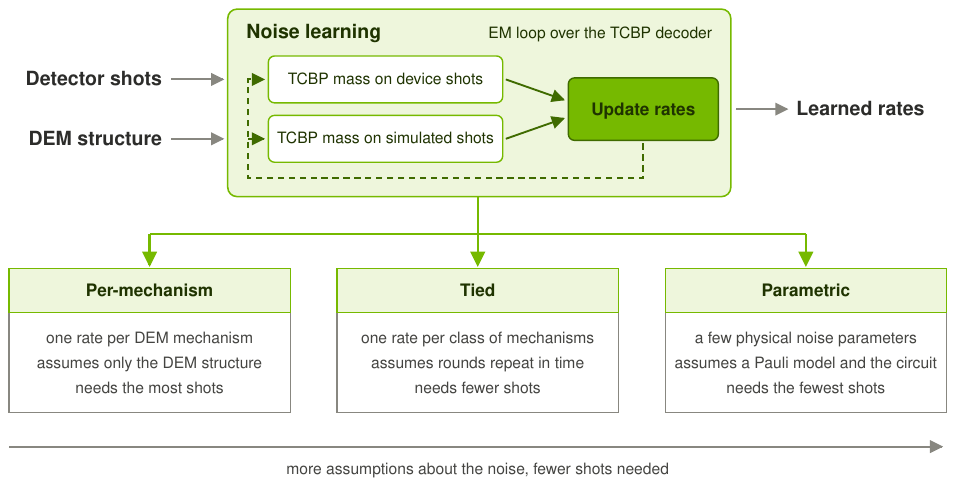}
\caption{\textbf{Noise learning with TCBP.} The inputs are recorded detector shots and the
structure of the detector error model; no error labels or nominal rates are used. Each
pass decodes the device shots and a reference set of shots simulated from the current
rates, sums the calibrated TCBP posterior of every mechanism over each set, and multiplies
each rate by the ratio of the two sums. The three variants share this loop and differ in
what is learned: one rate per mechanism, one rate per class of mechanisms that repeat
across rounds, or the parameters of a small physical noise model. Stronger assumptions
about the noise reduce the number of shots needed.}
\label{fig:noise-learning-overview}
\end{figure}

As detailed in Figure~\ref{fig:noise-learning-overview}, the input to the procedure
is a set of $N$ detector shots from the circuit under test. No error labels and no
nominal rates are used. The unknowns are the $m$ rates $p_j$ of the DEM, one per mechanism
($m = 67{,}752$ for the gross code below). The start is uniform: every $p_j$ is set to the
single rate that reproduces the mean detector event rate of the data.

Learning is an expectation-maximization (EM) loop over the TCBP decoder. The E-step decodes
the $N$ shots with the current rates and records, for every mechanism, the calibrated
posterior probability $\widehat p_j$ that TCBP assigns to it (Section~\ref{sec:tcbp}). Summing
over shots gives a posterior mass per mechanism. The same decoder with the same rates is
also run on a matched reference stream of $N$ shots simulated from the current rates. The
M-step multiplies each rate by the ratio of the two masses, device over reference. Where the
device data show more of a mechanism than the current model predicts, the rate goes up, and
vice versa. The reference stream cancels the residual bias of the TCBP calibration, because
the same miscalibration acts on both streams. The loop runs four passes, or fewer when the
rate update falls below a tolerance.

Two variants trade generality for shot efficiency. The tied variant shares one rate across
mechanisms that are translates of each other in time, which cuts the number of unknowns by
the number of rounds and reaches a given accuracy with fewer shots. The parametric variant
fits a small physical noise model, such as a per-gate Pauli channel, through a design
matrix that maps model parameters to DEM rates. Both reuse the same E-step. The results
below use the per-mechanism variant, which makes no assumption beyond the DEM structure.

\subsection{Experimental setup}

For experimental evaluation, we choose the $[[144,12,12]]$ bivariate-bicycle (gross) code, 12 rounds of
syndrome extraction, $Z$-basis memory, and circuit-level SI1000 noise at physical error rate
$p \in \{0.001, 0.002, 0.004\}$. The circuits and the seeded detector streams are those of
the Tesseract comparison. The DEM has 1{,}728 detectors, 67{,}752 mechanisms and 12
observables. All runs use one NVIDIA GB200.

The TCBP model is fitted once per $p$, in about 10\,s, on rates the fitting routine draws
itself, with no knowledge of the true rates. Learning data are the first $N$ shots of a late
segment of the stream, $N$ from $2^{10}$ to $2^{21}$, disjoint from the evaluation
shots. Learning time is wall-clock time of the EM loop on the GPU, excluding data loading
and the TCBP fit.

Decoding uses BP with 10 iterations in FP32, followed by JOBS with TCBP column ordering.
The learned rates are ordinary DEM priors and can be given to any decoder; only the EM
loop itself requires TCBP. We evaluate them with TCBP-JOBS because it is accurate and fast
enough for this experiment.
The same TCBP model is used for all four priors:
\emph{uniform}, the single start rate described above
($2.59 \times 10^{-4}$, $4.88 \times 10^{-4}$ and $8.71 \times 10^{-4}$ at the three $p$);
\emph{learned}, the per-mechanism rates after learning on $2^{21}$ shots with the default
stopping rule, which ends the EM loop after four passes here;
\emph{learned, 16 passes}, the rates after 16 passes with the tolerance off on $2^{19}$ shots
(905\,s on one GB200), the setting we recommend;
\emph{oracle}, the rates of the DEM the shots were sampled from.
Every arm decodes the same stream from shot 0 in chunks of 4{,}096. A shot fails if any of
the 12 observables is wrong or the correction does not reproduce the syndrome. The second
case did not occur. An arm stops at $2 \times 10^6$ shots or after 3{,}300\,s of wall time,
with no failure-based stop. Arms that hit the time cap decoded a prefix of the stream, so
they are compared shot by shot on the longest prefix all arms reached, with an exact McNemar
test on the shots where exactly one arm fails. Rate accuracy is the median over mechanisms
of $|\log(\widehat p_j / p_j)|$, where $p_j$ is the true rate. The same study was repeated
with a second seed for the TCBP fit and the learning run, and on the $X$-basis circuits.

\subsection{Results}

Table~\ref{tab:nl-learning} gives learning time and accuracy. Time is linear in $N$, about
0.43\,ms per shot, and does not depend on $p$. The start rate has median error 1.36, 1.30
and 1.18 at $p = 0.001$, $0.002$ and $0.004$. One run over $2^{17}$ shots takes 57\,s and
reduces it to 0.20 to 0.38. More shots help at $p = 0.002$ and $0.004$. At $p = 0.001$ the
error stops at 0.18 from $2^{19}$ shots on.

\begin{table}[t]
\centering
\caption{Noise learning on the gross code, one GB200. Error is the median over the 67{,}752
mechanisms of $|\log(\text{learned}/\text{true})|$. The uniform start has error
1.36 / 1.30 / 1.18 at $p = 0.001$ / $0.002$ / $0.004$. At $p = 0.001$ the $2^{21}$-shot run
stopped after three passes (682\,s).}
\label{tab:nl-learning}
\small
\begin{tabular}{rrrrr}
\toprule
 & & \multicolumn{3}{c}{median $|\log(\text{learned}/\text{true})|$} \\
\cmidrule(lr){3-5}
$N$ & time & $p{=}0.001$ & $0.002$ & $0.004$ \\
\midrule
$2^{17}$ & 57\,s  & 0.252 & 0.200 & 0.377 \\
$2^{19}$ & 227\,s & 0.179 & 0.145 & 0.339 \\
$2^{21}$ & 910\,s & 0.182 & 0.119 & 0.326 \\
\bottomrule
\end{tabular}
\end{table}

Table~\ref{tab:nl-ler} gives the logical error rate (LER) of TCBP-JOBS under the four
priors. Intervals are 95\% Clopper--Pearson. Learning from the uniform start lowers the LER
by a factor of 22 at $p = 0.002$ and 6 at $p = 0.004$ with the default stopping rule, and by
27 and 9 with 16 passes. It also makes decoding 3 to 7 times faster per shot, because BP
under the uniform prior converges on far fewer shots and JOBS runs more often. Against the
oracle, the learned prior with the default stopping rule is 1.2 times worse at $p = 0.002$
(81 against 66 failures; on the common prefix 4 shots fail only under learned and 2 only
under oracle) and 1.45 times worse at $p = 0.004$, where the paired test on the common
prefix of 28{,}672 shots is decisive: 1{,}303 shots fail only under learned and 477 only
under oracle. With 16 passes the gap closes: 1.03 at $p = 0.002$ (68 against 66 failures,
13 against 11 discordant, $p = 0.84$) and 1.02 at $p = 0.004$ (3{,}143 against 2{,}907
discordant on the common prefix of 172{,}032 shots, $p = 3 \times 10^{-3}$, significant only
because that prefix holds 11{,}000 failures). Against the default-rule learned prior the 16-pass prior is 0.84 times
at $p = 0.002$ ($p = 0.06$) and 0.71 times at $p = 0.004$ ($p < 10^{-300}$). The $X$-basis
circuits give the same picture: default rule 64 against 50 oracle failures at $p = 0.002$
and 1.45 at $p = 0.004$; 16 passes 50 against 50 (1.00) and 1.03. Learning on $2^{17}$
shots with 16 passes (227\,s) is already within 1.00 to 1.08 of the oracle, and learning on
$2^{21}$ shots (3{,}620\,s) halves the rate error again, to 0.05, but moves the LER by at
most one discordant shot in $2 \times 10^6$. We therefore recommend the learned prior as 16
EM passes with the tolerance off on $2^{19}$ shots, 905\,s on one GB200, within 1.02 to 1.03
of the DEM prior at $p = 0.002$ to $0.004$. At $p = 0.001$ every learned and oracle prior fails 0 or 1 shots in
$2 \times 10^6$ (the one $X$-basis failure is the same shot under the oracle and both
learned priors), so no claim is made there beyond the 9-to-1 drop from the uniform prior.
The second seed changes every failure count by 1 to 2\%. The same picture was obtained on
an independent SI1000 construction of the gross code, the unmodified $Z$-memory circuit of the
Tesseract repository's test data, with freshly sampled shots at $p = 0.002$, $0.003$ and
$0.004$ and every arm stopped at 200 failures, which fixes each ratio to about 20\%: the
learned prior is at 1.05, 1.00 and 1.03 times the oracle LER with 16 passes and at 1.06, 1.20
and 1.39 times with the default rule.

\begin{table*}[t]
\centering
\caption{Logical error rate of TCBP-JOBS on the gross code under four rate priors, $Z$
basis, one seed. Uniform: one rate matched to the mean detector event rate. Learned:
per-mechanism rates from $2^{21}$ unlabeled shots with the default stopping rule (four
passes). Learned, 16 passes: from $2^{19}$ shots with 16 passes and the tolerance off.
Oracle: the true rates. Every arm decodes
the same stream from shot 0. Arms with fewer than $2 \times 10^6$ shots stopped at the
3{,}300\,s wall-time cap. Decode time is per shot, batched, on one GB200, measured
separately with CUDA-Q QEC 0.9.0 on the first 131{,}072 shots of each arm (28{,}672 under the
uniform prior at $p = 0.004$).}
\label{tab:nl-ler}
\small
\begin{tabular}{llrrlr}
\toprule
$p$ & prior & shots & fails & LER (95\% CI) & $\mu$s/shot \\
\midrule
0.001 & uniform & 2{,}000{,}000 & 9 & $4.5 \times 10^{-6}$ ($2.1$--$8.5 \times 10^{-6}$) & 399 \\
0.001 & learned & 2{,}000{,}000 & 1 & $5 \times 10^{-7}$ ($0.01$--$2.8 \times 10^{-6}$) & 150 \\
0.001 & learned, 16 passes & 2{,}000{,}000 & 0 & $0$ ($0$--$1.8 \times 10^{-6}$) & 149 \\
0.001 & oracle  & 2{,}000{,}000 & 0 & $0$ ($0$--$1.8 \times 10^{-6}$) & 149 \\
\midrule
0.002 & uniform & 315{,}392 & 286 & $9.1 \times 10^{-4}$ ($8.1$--$10.2 \times 10^{-4}$) & 3{,}007 \\
0.002 & learned & 2{,}000{,}000 & 81 & $4.1 \times 10^{-5}$ ($3.2$--$5.0 \times 10^{-5}$) & 433 \\
0.002 & learned, 16 passes & 2{,}000{,}000 & 68 & $3.4 \times 10^{-5}$ ($2.6$--$4.3 \times 10^{-5}$) & 413 \\
0.002 & oracle  & 2{,}000{,}000 & 66 & $3.3 \times 10^{-5}$ ($2.6$--$4.2 \times 10^{-5}$) & 408 \\
\midrule
0.004 & uniform & 28{,}672 & 16{,}441 & $0.573$ ($0.568$--$0.579$) & 20{,}251 \\
0.004 & learned & 131{,}072 & 12{,}260 & $9.4 \times 10^{-2}$ ($9.2$--$9.5 \times 10^{-2}$) & 5{,}623 \\
0.004 & learned, 16 passes & 172{,}032 & 11{,}347 & $6.6 \times 10^{-2}$ ($6.5$--$6.7 \times 10^{-2}$) & 4{,}354 \\
0.004 & oracle  & 180{,}224 & 11{,}641 & $6.5 \times 10^{-2}$ ($6.4$--$6.6 \times 10^{-2}$) & 4{,}158 \\
\bottomrule
\end{tabular}
\end{table*}

\begin{figure}[t]
\centering
\includegraphics[width=0.85\linewidth]{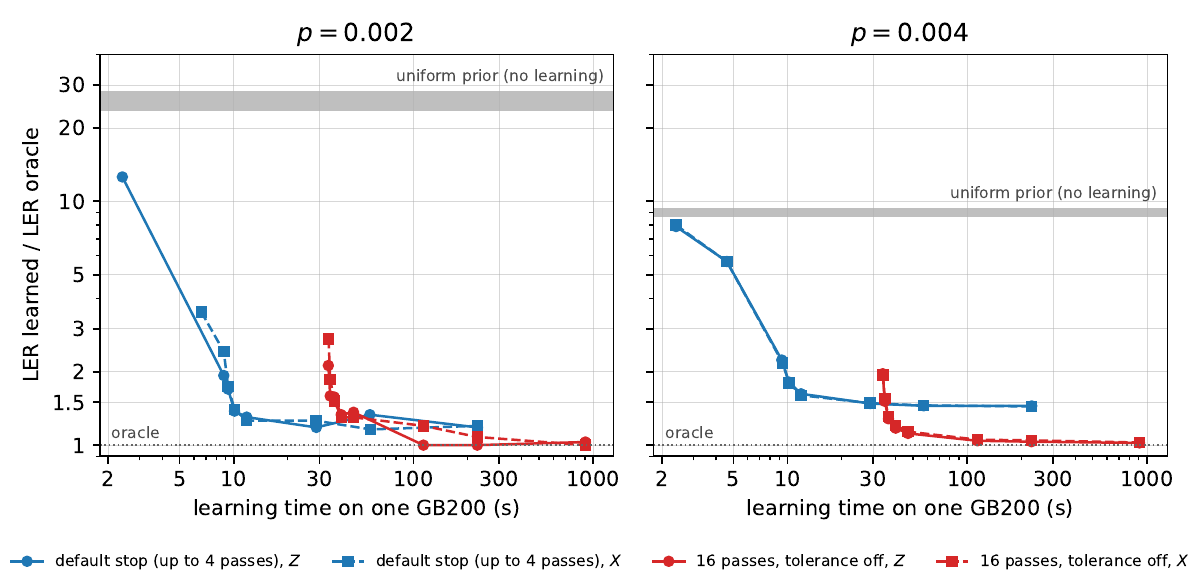}
\caption{Learned over oracle logical error rate on the common prefix against the learning
time on one GB200, gross code, one panel per $p$; the learning shots run from $2^{10}$ to
$2^{19}$ along each curve. Blue: default stopping rule. Red: 16 passes with the tolerance
off. Circles and solid lines: $Z$ memory; squares and dashed lines: $X$ memory. The gray
band is the uniform prior, that is, no learning; the dotted line is the oracle. The learning
time excludes the TCBP model fit, about 10\,s. Arms with a poor prior decode more slowly and
reach the 3{,}300\,s cap on a shorter prefix. At $p = 0.001$ every learned arm fails 0 or 1 shots
in $2 \times 10^6$ and is not shown.}
\label{fig:nl-ler-vs-time}
\end{figure}

\paragraph{Stopping rule versus shot count.}
The decoder's accuracy is set by when the EM loop stops rather than by how many shots it sees.
Table~\ref{tab:nl-stop} and Figure~\ref{fig:nl-ler-vs-time} compare the default rule, which
stops after four passes from $2^{12}$ shots up, with 16 passes and the tolerance off, each
learned from $N = 2^{10}$ to $2^{19}$ shots and then decoded over the same $2 \times 10^6$
shots as the oracle arm; the ratio is learned failures over oracle failures on the common
prefix. From $2^{14}$ shots up the ratio under the default rule is flat in $N$: 1.2 to 1.3
at $p = 0.002$ and 1.45
to 1.6 at $p = 0.004$, up to the $2^{21}$-shot run of
Table~\ref{tab:nl-ler}, although the median rate error keeps falling, from 0.33 to 0.12 at
$p = 0.002$. With 16 passes the arm reaches the oracle at $p = 0.002$ from $2^{16}$ shots
in $Z$ memory (ratio 1.00, 18 shots fail only under learned against 18 only under oracle,
$p = 1.00$; the same at $2^{17}$, 15 against 15), and in $X$ memory from $2^{17}$ shots
(1.08, $p = 0.50$, then 1.00 at $2^{19}$). At $p = 0.004$ it reaches 1.02 to 1.05 from
$2^{16}$ shots in both bases; the 1.02 at $2^{19}$ is still significant against the oracle
($p = 3 \times 10^{-3}$) because the common prefix holds 11{,}000 failures. Learning time is
linear in shots times passes: 12, 29, 57 and 228\,s under the default rule and 46, 114,
227 and 905\,s with 16 passes at $2^{14}$, $2^{16}$, $2^{17}$ and $2^{19}$ shots on one
GB200. The cheapest prior indistinguishable from the oracle at $p = 0.002$ is therefore 16
passes on $2^{16}$ shots, 114\,s, against 909\,s for the default $2^{21}$-shot run that stays
at 1.23. Below $2^{16}$ shots the 16-pass prior no longer reaches the oracle, because it is
then limited by the shots and not by the passes.

\paragraph{Accuracy versus learning time.}
Figure~\ref{fig:nl-ler-vs-time} plots the same ratio against the wall time of the learning
run. Without learning, the uniform prior is 24 to
27 times worse than the oracle at $p = 0.002$ and 9 times worse at $p = 0.004$. Under the
default rule the learner makes a single pass over $2^{10}$ shots in 2.4\,s, and the prior is
still 8 to 13 times worse than the oracle, with a median rate error of 1.1 (the $X$-memory
run at $p = 0.002$ made three passes, 6.6\,s, and reached 3.5). Four passes over $2^{12}$ or
$2^{13}$ shots take 9 to 10\,s and bring the ratio to 1.4 to 1.7 at $p = 0.002$ and 1.8 to
2.2 at $p = 0.004$; $2^{14}$ shots, 12\,s, give 1.3 and 1.6. Most of the gain over the
uniform prior is therefore obtained in about 10\,s of learning on one GB200. The remaining
factor of 1.2 to 1.6 does not shrink with more shots under the default rule and is removed
only by the 16-pass setting on $2^{16}$ shots or more, from 114\,s. Below $2^{14}$ shots the
learning time is no longer proportional to $N$, since a pass costs about 2\,s whatever the
shot count; 16 passes then take 34 to 40\,s, and their prior is limited by the shots: its
median rate error stays at 0.45 to 0.65 from $2^{10}$ to $2^{13}$ shots and the ratio at 1.2
to 2.7. Every prior learned from fewer than $2^{14}$ shots is significantly worse than the
oracle at the 5\% level. The times exclude the TCBP model fit, about 10\,s per $p$, which
every prior needs, the oracle included.

\begin{table}[t]
\centering
\caption{Learned over oracle logical error rate on the common prefix against learning shots
$N$ and EM stopping rule, gross code, $Z$ memory, seed 1. Default: the default stopping rule, which runs
four passes from $2^{12}$ shots up, one pass at $2^{10}$, and at $2^{11}$ four passes at
$p = 0.002$ and two (4.6\,s) at $p = 0.004$.
16 passes: tolerance off. Time is the learning time on one GB200, default / 16 passes. The
$2^{21}$ row is the learned arm of Table~\ref{tab:nl-ler} under the default rule and the
$2^{21}$-shot 16-pass learning. From $2^{14}$ shots up $X$ memory
differs by at most 0.08 except at $2^{17}$ under the default rule (1.16) and at $2^{16}$
with 16 passes (1.20, $p = 0.08$); below $2^{14}$ shots it is shown in
Figure~\ref{fig:nl-ler-vs-time}.}
\label{tab:nl-stop}
\small
\begin{tabular}{rrrrrr}
\toprule
 & & \multicolumn{2}{c}{$p = 0.002$} & \multicolumn{2}{c}{$p = 0.004$} \\
\cmidrule(lr){3-4} \cmidrule(lr){5-6}
$N$ & time (s) & default & 16 passes & default & 16 passes \\
\midrule
$2^{10}$ & 2.4 / 34  & 12.6 & 2.12 & 7.88 & 1.96 \\
$2^{11}$ & 8.8 / 34  & 1.93 & 1.59 & 5.67 & 1.51 \\
$2^{12}$ & 9.3 / 36  & 1.70 & 1.58 & 2.23 & 1.28 \\
$2^{13}$ & 10 / 40   & 1.38 & 1.33 & 1.81 & 1.17 \\
$2^{14}$ & 12 / 46   & 1.30 & 1.36 & 1.62 & 1.12 \\
$2^{16}$ & 29 / 114  & 1.18 & 1.00 & 1.48 & 1.04 \\
$2^{17}$ & 57 / 227  & 1.33 & 1.00 & 1.45 & 1.03 \\
$2^{19}$ & 228 / 905 & 1.18 & 1.03 & 1.45 & 1.02 \\
$2^{21}$ & 909 / 3623 & 1.23 & 1.03 & 1.44 & 1.02 \\
\bottomrule
\end{tabular}
\end{table}

\subsection{Discussion}

At $p = 0.002$ a learning run over $2^{17}$ shots costs about the same GPU time as decoding
$1.3 \times 10^5$ shots with the learned prior, or $1.9 \times 10^4$ shots with the uniform
prior. At this cost the rates can be re-learned between calibrations, from
the shots the experiment produces anyway, without labels and without stopping the device. The
decoder then keeps a current rate model instead of the one from the last calibration. Where
the decoder cannot wait that long, four passes over $2^{13}$ shots give a prior within a
factor of 1.4 to 1.8 of the oracle in 10\,s (Figure~\ref{fig:nl-ler-vs-time}). A warm
start from the previous rates needs fewer passes than the uniform start used here. The tied
and parametric variants reach the same accuracy from fewer shots, which shortens the window
over which the noise must be stationary.

\subsection{Limitations}

The learning data are simulated from the same DEM the oracle uses, so the study measures
recovery of a known model, not model mismatch on hardware. At $p = 0.001$ the learned and
oracle arms have 0 or 1 failures in $2 \times 10^6$ shots, so their LERs are upper bounds of
about $2 \times 10^{-6}$, not measurements. At $p = 0.001$ learning on $2^{21}$ shots ran
three passes and did not improve on $2^{19}$ shots. At $p = 0.004$ the arms are time-capped
at unequal shot counts, and the paired comparison on the common prefix is the equal-shot
result. The second seed varies only the TCBP fit and the learning run, since the shots are
fixed. The median log error can hide a tail of badly estimated mechanisms that matter for
decoding. The uniform prior is the learner's data-derived start rate, one value for all
mechanisms. A different uniform choice, such as the nominal $p$, would change the uniform
arm but not the learned or oracle arms.

%% file: inputs/06complimentary-gap.tex
\section{Complementary gap and magic-state cultivation}
\label{sec:complementary-gap}

TCBP-JOBS naturally provides a complementary-gap score through its decoding
search. We first evaluate this score as a postselection criterion in memory
experiments, then apply it to magic-state cultivation. In cultivation, TCBP-JOBS reduces the binary-$Y$ diagnostic
error---the fraction of accepted shots whose decoder-corrected
logical-$Y$ expectation has the wrong sign relative to the
target state~\cite{chase2026clifft}---by up to $4.7\times$
relative to the original desaturation/matching decoder used by Gidney et al.~\cite{gidney2024cultivation},
while using the same cultivation and grafted surface-escape
circuit and hard-postselection rules.
Combining JOBS with direct color-code escape yields up to
$2.8\times$ higher modeled throughput per installed qubit
than the original surface-matching configuration at the
selected binary-$Y$ error ceilings.

The complementary gap measures the cost separation between competing logical
explanations of the observed syndrome. Let $H$ map modeled faults $e$ to detector
outcomes and $L$ map them to logical labels. In the decoder's independent-binary
fault model, with fault probabilities $p_j$, the minimum cost in class $\ell$
and the exact complementary gap are
\begin{align}
 W_\ell(s)&=\min_{He=s,\,Le=\ell}
       \sum_j e_j\log\frac{1-p_j}{p_j},\\
 \Delta(s)&=\min_{\ell\ne\ell_*}W_\ell(s)-W_{\ell_*}(s),
 \label{eq:cult-gap}
\end{align}
where $s$ is the syndrome and $\ell_*$ minimizes $W_\ell$.

TCBP guides search ordering. JOBS records candidate costs and logical labels,
including for detector error models with hyperedges. The score used here,
$\widehat\Delta_{\rm JOBS}$, is the difference between the lowest encountered
costs in the two best logical classes. The correction returned by JOBS is
selected separately and need not belong to the lowest-cost encountered class.
These minimum costs do not sum probabilities over all faults within a class,
and finite search may miss competing classes. We therefore evaluate the score
empirically through observed error rates after postselection.
When only one class is encountered, the score is infinite. These shots are
retained, although some still contain errors.

\subsection{Postselection on memory experiments}
\label{sec:gap-memory}
A decoder that returns only a correction gives no indication of how close the decision
was. The score $\widehat\Delta_{\rm JOBS}$ measures ambiguity among the logical
classes encountered during search. A small score indicates competing classes
with similar candidate costs, while a large score indicates greater
separation among the encountered classes. We test whether discarding
low-score shots preferentially removes decoding failures.

\begin{figure*}[!tbp]
 \centering
 \includegraphics[width=\textwidth]{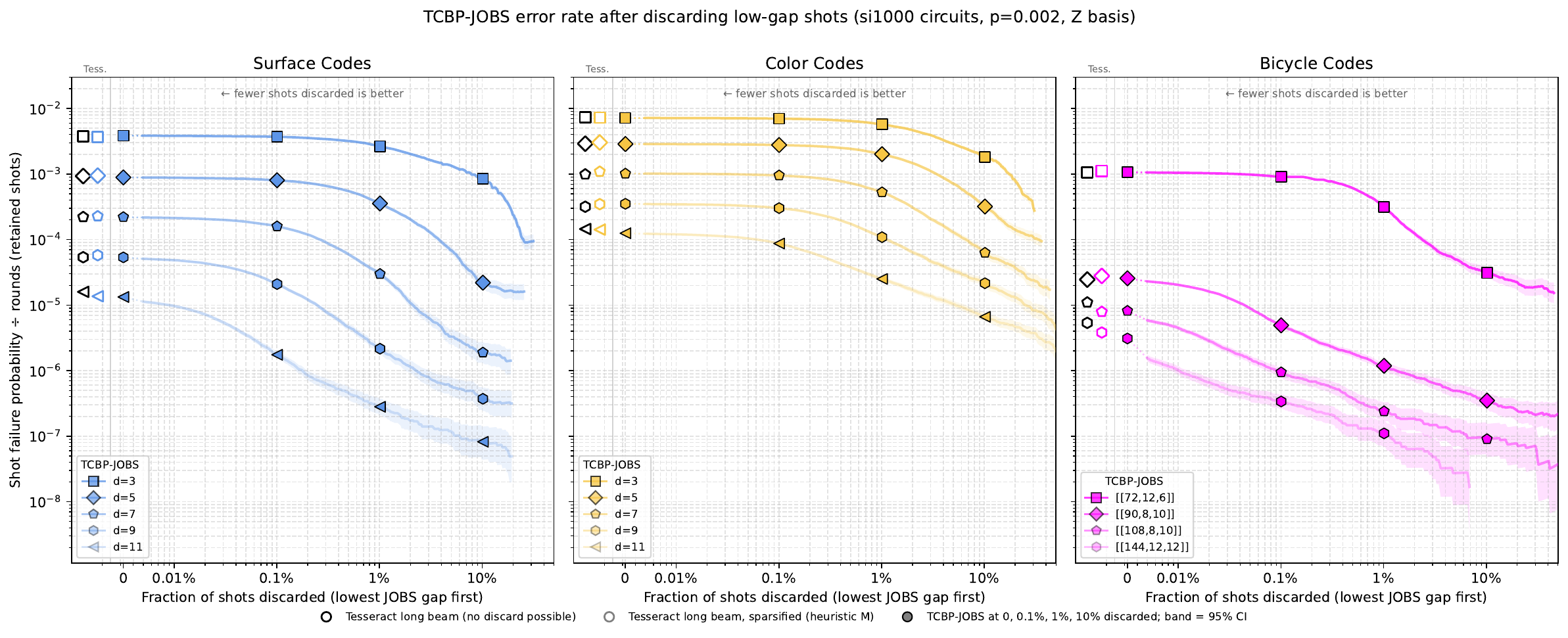}
 \caption{\textbf{Postselection on the JOBS gap in memory experiments.}
 Failure probability per round of the retained shots against the fraction of shots
 discarded, lowest $\widehat\Delta_{\rm JOBS}$ first, for SI1000 circuits at $p=0.002$ in the $Z$ basis.
 Filled markers show TCBP-JOBS at 0, $0.1\%$, $1\%$ and $10\%$ discarded; bands are $95\%$
 confidence intervals. Open markers show long-beam Tesseract, with and without
 sparsification; the evaluated implementation provides no comparable postselection score.}
 \label{fig:jobs-gap-discard}
\end{figure*}

Figure~\ref{fig:jobs-gap-discard} tests this on memory circuits of the three code families
of Section~\ref{sec:jobs-vs-tesseract} under SI1000 noise at $p=0.002$ in the $Z$ basis. JOBS searches
every shot, including those on which BP converged, so that every shot receives a score.
We sort the shots by $\widehat\Delta_{\rm JOBS}$ and discard those with the lowest score
first; shots in which only one class was encountered have an infinite score and are
discarded last. The figure plots the failure probability per round of the retained shots
against the fraction discarded. Long-beam Tesseract is shown for reference at
zero discard; the evaluated implementation does not provide a comparable postselection score.

\begin{figure*}[!tp]
 \centering
 \includegraphics[width=\textwidth]{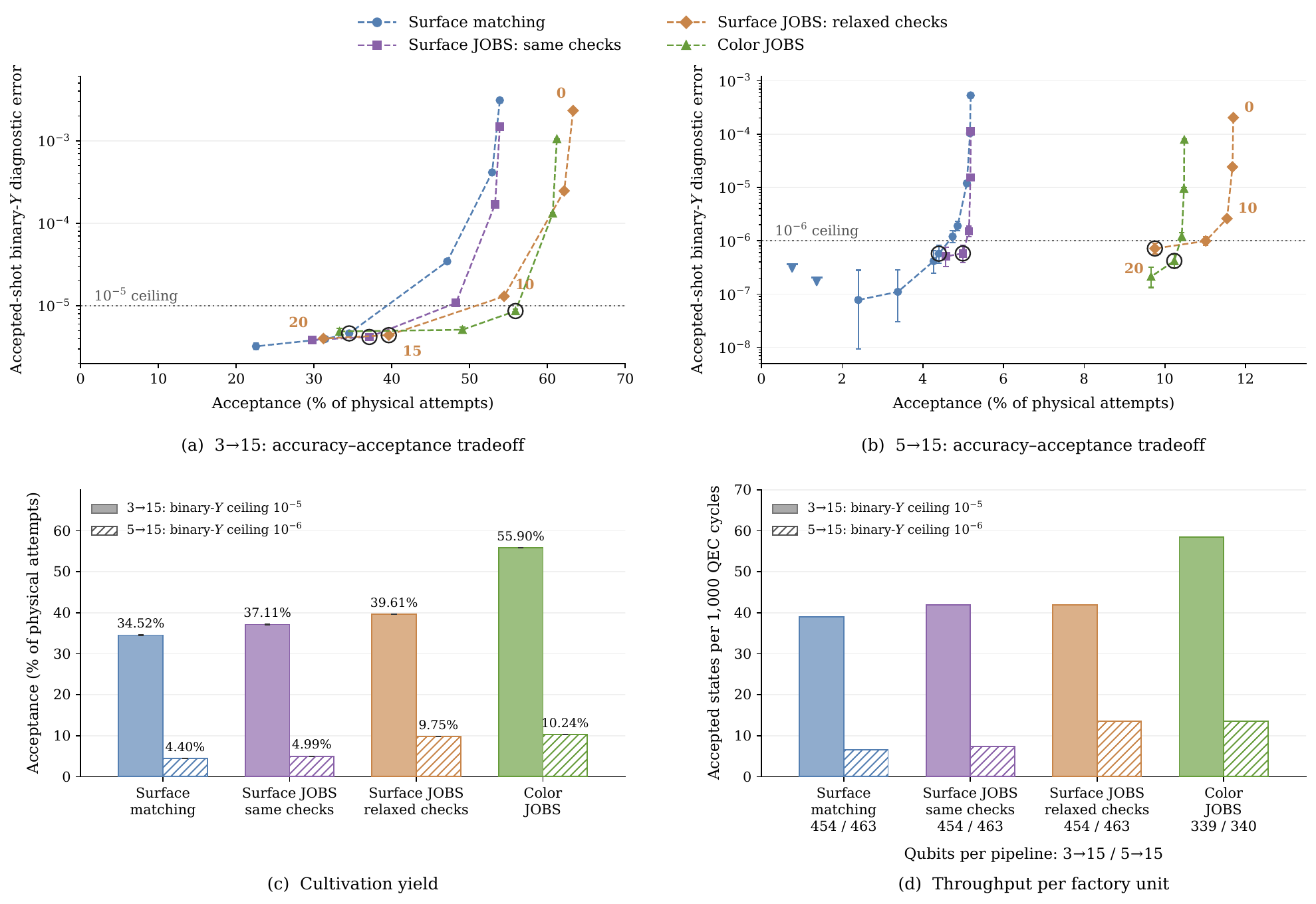}
 \caption{\textbf{Complementary-gap postselection, cultivation yield and throughput.}
 (a,b) Accepted-shot binary-$Y$ diagnostic error versus acceptance for
 $3\!\to\!15$ and $5\!\to\!15$ at circuit-level noise $p=10^{-3}$.
 Markers denote tested gap thresholds; dashed lines are visual guides.
 Error bars are pointwise two-sided $95\%$ Clopper--Pearson intervals.
 Downward arrows show one-sided $95\%$ upper bounds where no errors were observed.
 Orange labels give selected relaxed-JOBS thresholds in nats.
 Black rings identify the operating points used in (c,d).
 (c) Highest measured acceptance among tested thresholds whose two-sided
 $95\%$ error interval has its upper endpoint below $10^{-5}$ for
 $3\!\to\!15$ (solid bars) or $10^{-6}$ for $5\!\to\!15$ (hatched bars).
 Acceptance error bars are pointwise $95\%$ intervals. Intervals are not adjusted
 for selection among thresholds.
 (d) Accepted states per 1,000 reference QEC cycles per factory unit, including
 rejected work with early abort. One reference cycle is assigned ten
 equal-duration circuit steps, matching the repeated escape syndrome-extraction
 rounds in both constructions. This is an approximate time unit, not a count
 of syndrome rounds experienced by each state.
 Footprints below the bars give physical qubits per unit for $3\!\to\!15$ / $5\!\to\!15$.}
 \label{fig:cultivation-gap}
\end{figure*}

\subsection{Cultivation experiment}
Gidney et al.\ combine cultivation and complementary-gap postselection,
injecting and cultivating in a color code before growing into a grafted,
matchable code~\cite{gidney2024cultivation}. We compare their desaturation/matching
decoder with TCBP-JOBS on that construction, and with direct growth in triangular
6.6.6 color codes. We call the grafted route ``surface'' for brevity.
The label $d_1\!\to\!15$ denotes cultivation and final code distances.

We simulate non-Clifford $T/T^\dagger$ gates using CliffT~\cite{chase2026clifft},
with uniform circuit-level Pauli noise $p=10^{-3}$. For $d_1=3$, we simulate
$2^{28}$ color attempts and $5\cdot2^{27}$ surface attempts. For $d_1=5$,
we simulate $2^{30}$ attempts for each construction. JOBS uses 16 search draws.
The three surface policies use the same simulated shots. The JOBS ``same checks'' policy retains the original matching baseline's hard-postselection rules. The ``relaxed checks'' policy removes automatic rejection on selected escape-stage detector events, while retaining their outcomes for decoding and preserving cultivation postselection.
The direct color-code escape experiment uses JOBS on a different code construction than that of Gidney et al.~\cite{gidney2024cultivation} and therefore tests the combined effect of
the construction and decoding choices.

Following the released end-to-end analysis accompanying
CliffT~\cite{chase2026clifft}, we classify each simulated shot
using its decoder-corrected logical-$Y$ expectation value,
evaluated before terminal stabilizer projection.
In the target frame, where the ideal state has
$\langle Y_L\rangle=1/\sqrt{2}$, define
\begin{equation}
\begin{aligned}
 D_Y &= \frac{1}{2}
       +\frac{\langle Y_L\rangle_{\rm corr}}{\sqrt{2}},\\
 p_{\rm binY} &= \Pr(D_Y<1/2\mid\mathrm{accepted}).
\end{aligned}
\end{equation}
We call $p_{\rm binY}$ the binary-$Y$ diagnostic error rate
and estimate it as the fraction of accepted shots satisfying
$D_Y<1/2$, equivalently $\langle Y_L\rangle_{\rm corr}<0$.
This criterion thresholds a simulated expectation value,
not an individual $Y_L$ measurement outcome, and is distinct
from full magic-state infidelity. Evaluating full magic-state infidelity would also require a
full-Pauli recovery for matching, whose analysis specifies only
the correction needed for this diagnostic. Acceptance is the fraction of all physical attempts retained,
including hard rejects in the denominator.

\subsection{Accuracy, postselection and factory resources}
With identical hard postselection and no score cut, TCBP-JOBS reduces the
binary-$Y$ diagnostic error by $2.1\times$ for $3\!\to\!15$ and $4.7\times$
for $5\!\to\!15$, without changing the quantum circuit or accepted population.
The respective error rates decrease from $3.09\times10^{-3}$ to
$1.48\times10^{-3}$ and from $5.31\times10^{-4}$ to $1.13\times10^{-4}$.
For $3\!\to\!15$, applying a JOBS score cut of 15 nats further lowers the
error rate to $4.19\times10^{-6}$, while acceptance falls from $53.88\%$
to $37.11\%$. These comparisons test the complete decoder.
Isolating TCBP's contribution requires an ablation.

Panels (a,b) of Figure~\ref{fig:cultivation-gap} show how complementary-gap thresholds
trade acceptance for diagnostic accuracy in cultivation. The same-check surface
curves compare the two decoders on a common hard-selected population; relaxed
surface checks and direct color escape provide additional operating points.
The two downward arrows in (b) are surface-matching thresholds with zero
observed errors; they indicate statistical upper bounds, not zero error rates.

Figure~\ref{fig:cultivation-gap}(c) compares exploratory yields. For each policy,
we select the highest acceptance among tested thresholds whose $95\%$ binary-$Y$
error interval has its upper endpoint below the target.
With identical hard checks, JOBS raises acceptance from $34.52\%$ to $37.11\%$
for $3\!\to\!15$, and from $4.40\%$ to $4.99\%$ for $5\!\to\!15$.
Relaxing escape checks further raises the latter yield to $9.75\%$.
This gain comes from changing the rejection policy. Matching with equivalently
relaxed checks remains untested, leaving open whether it offers the same benefit.
Direct color escape reaches $10.24\%$. The four selected $5\!\to\!15$ policies,
in that order, contain 27, 31, 75 and 46 binary errors, with different measured
error rates. These binary-$Y$ error ceilings do not establish corresponding
bounds on full magic-state infidelity.

A \emph{factory unit} is one reusable injection--cultivation--escape pipeline,
attempting one state at a time. Its accepted rate is $r=a/\bar\tau$, where $a$
is acceptance and $\bar\tau$ is mean time per attempt. We assign equal duration
to the simulator's circuit steps (Stim intervals) and define one reference
QEC-cycle duration as ten such steps. This matches the repeated escape rounds
in both constructions: one reset layer, eight CNOT layers and one measurement
layer. We use this duration as an approximate time unit throughout the factory,
including injection and cultivation; it does not count actual syndrome rounds
or calibrate unequal hardware operation durations.
We assume an attempt stops at the end of
the step containing its first rejecting check. Attempts rejected by the decoder's
gap consume the full circuit time.

Figure~\ref{fig:cultivation-gap}(d) shows the accepted-state production rate
per factory unit. Surface matching and direct color JOBS produce respectively
39.0 and 58.4 accepted states per 1,000 reference QEC cycles for $3\!\to\!15$,
and 6.51 and 13.6 for $5\!\to\!15$. Combining throughput and footprint,
direct color escape with JOBS provides $2.0\times$ and $2.8\times$ higher
modeled throughput per installed qubit than surface matching for
$3\!\to\!15$ and $5\!\to\!15$, respectively, at the selected binary-$Y$
error ceilings. These gains combine architecture, decoder and
rejection-policy changes. At $5\!\to\!15$, direct color and relaxed
surface JOBS have nearly equal throughput per unit, while color uses
340 versus 463 qubits. At $3\!\to\!15$, relaxing surface checks increases
yield but barely changes throughput because more time is spent on
attempts that would otherwise stop early.

Closely spaced $T$ gates may require parallel factory units or states prepared
and stored in advance. Hardware estimates must also include gate durations,
decoding and transfer times, and errors accumulated during storage.

%% file: inputs/07importance-sampling.tex
\section{Importance Sampling}
\label{sec:importance-sampling}

The logical error rates shown in Section~\ref{sec:jobs} extend down to $p=0.001$, where TCBP-JOBS fails on 55 of approximately $2.2\times 10^8$ shots of the $[[144,12,12]]$ gross code, including both $X$ and $Z$ memory experiments. Probing the logical error rate at significantly lower physical error rate is prohibitively expensive with Monte Carlo. At a logical error rate of order $10^{-12}$, direct Monte Carlo sampling requires approximately $10^{14}$ shots to observe 100 failures. In this section we use the splitting method of Bravyi and Vargo~\cite{bravyi2013simulation} to explore this regime. Splitting expresses the logical error rate at low $p$ as the directly measured rate at high $p$ multiplied by a product of failure-probability ratios between neighboring values of $p$. Each ratio is an average of likelihood ratios over fault configurations on which the decoder fails, sampled by a Markov chain (Appendix~\ref{app:splitting}). 

We study the gross code's 12-round $Z$ memory experiment under SI1000 noise, from $p=0.004$ to $p=10^{-4}$. The method requires one decoder along a run, so each $p$ for which we estimate the logical error is associated with one full splitting run from $p_0$ to $p$ using a decoder trained at $p$. 

\begin{figure}[t]

\centering
\includegraphics[width=\linewidth]{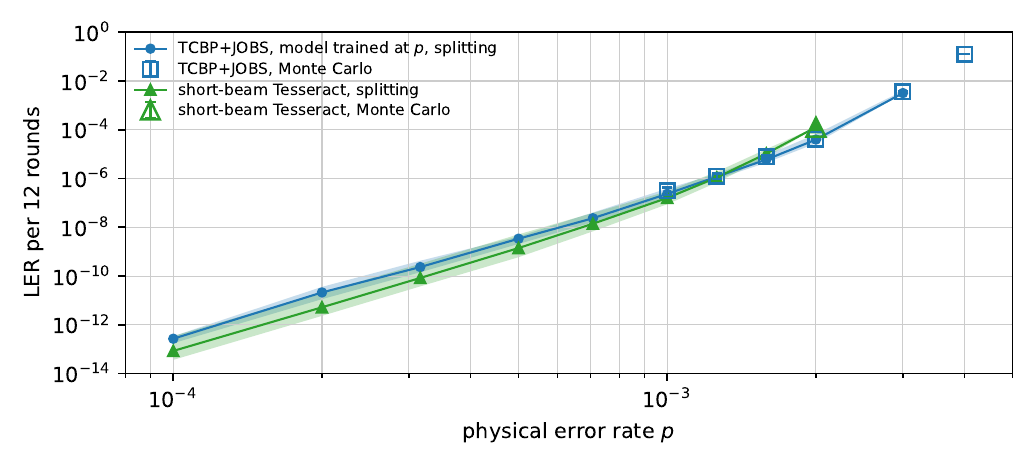}
\caption{Logical error rate per 12 rounds of the $[[144,12,12]]$ code, $Z$ memory, SI1000 noise. Filled symbols: splitting, median of 8 independent runs, bands show the range of the runs. Open symbols: Monte Carlo by the same decoder, with 95\% intervals. Circles and squares: TCBP-JOBS with a TCBP model trained at the DEM rates of each $p$. Every splitting point from $p = 3 \times 10^{-3}$ to $10^{-4}$ is the endpoint of its own set of runs, and the Monte Carlo points from $p = 4 \times 10^{-3}$ to $10^{-3}$ use the model trained at the sampled $p$. Triangles: short-beam Tesseract, same prior at all $p$, with splitting started from the Monte Carlo point at $p = 0.002$.}
\label{fig:splitting-ler}
\end{figure}

Where splitting and Monte Carlo methods overlap, they produce consistent logical error rate estimates using the same decoder. We have no independent verification below the range accessible to direct Monte Carlo sampling, and incomplete exploration of the failure set can potentially bias the splitting estimates downward. We use 128 parallel chains per run to improve exploration, and 8 independent runs to assess statistical variation. Fig.~\ref{fig:splitting-ler} shows the median and range of these estimates. This range is not a confidence interval, though, and cannot indicate any bias shared across all 8 runs. 

The low-error-rate regime reveals decoder limitations that are difficult to probe with direct sampling. In particular, failure to correct low-weight fault configurations can significantly increase the logical error rate. Splitting samples these rare failures and can therefore provide examples that help diagnose decoder limitations. For a fixed decoder under SI1000 noise, each fault probability scales linearly with $p$ as $p\to0$, so the asymptotic logarithmic slope equals the minimum number of faults that causes a decoder failure. For an optimal decoder of the gross code's standard circuit, this exponent is 5, corresponding to its circuit-level distance of 10~\cite{cheng2026certifying}.

In a separate experiment using the original TCBP model of Section~\ref{sec:tcbp}, the measured local slope decreases to approximately 5 at $p=10^{-4}$. However, this is not its asymptotic limit. Among 3,840 stored failing configurations at $p=2\times10^{-4}$, two contain four faults and 16 contain five. The presence of four-fault failures establishes that the asymptotic exponent of this decoder configuration is at most 4. Modifications to the decoder parameters (in particular increasing the JOBS draws from 16 to 64), or using Tesseract with the same prior, corrects all 18 configurations, showing that these particular failures can potentially be avoided. 

The TCBP-JOBS experiments used 8 independent splitting runs, requiring $2\times10^7$ decoder calls and just over 1 node-hour per run on nodes equipped with four B200 GPUs. Increasing the JOBS draws from 16 to 64 increased the cost per run to approximately 2.5 node-hours. We estimate that a direct Monte Carlo calculation of the TCBP-JOBS LER using approximately $10^{14}$ shots at $p=10^{-4}$ would require about $5\times10^6$ node-hours, or approximately 600 node-years. The eight short-beam Tesseract runs required approximately 24 hours on a single 144-core CPU.

%% file: inputs/08conclusions.tex
\section{Conclusions}
\label{sec:conclusions}

We introduced trajectory-calibrated belief propagation and jittered OSD beam search as a GPU-accelerated approach to high-accuracy quantum error correction decoding.
Across 72 surface, color and bivariate-bicycle memory circuits at $p=0.1\%$ to $0.4\%$, TCBP-JOBS achieves logical error rates comparable to or lower than the tested Tesseract configurations in nearly all settings, with the largest gains on
the largest codes at $p \ge 0.2\%$. Serial decoding is $10\times$ to $7{,}400\times$ faster than sparsified Tesseract and up to $13{,}000\times$ faster than its unsparsified long-beam preset, and one GPU decodes the 24.2-million-shot
Tesseract benchmark in 45 minutes. Mean serial latencies stay below 18\,ms per memory shot and are 13.5\,ms per gross-code surgery shot at $p=0.2\%$, which is relevant to AMO feed-forward workloads.
The same components serve three further tasks. TCBP's calibrated probabilities support unsupervised noise learning that comes within 2--3\% of the oracle-prior logical error rate on simulated gross-code data. The JOBS search provides a
complementary-gap score for postselection in memory experiments and magic-state cultivation. Its speed makes rare-event splitting affordable down to logical error rates near $10^{-12}$; at the physical error rates this reaches ($10^{-4}$ to $5\times10^{-4}$), Tesseract is more accurate than TCBP-JOBS on the gross code.

\paragraph{Future directions.}
With minor exceptions\footnote{Some experiments require enabling options to extract
supplementary outputs like complementary gap or control whether or not to perform
JOBS searches on BP-converged shots.} \footnote{Section~\ref{sec:importance-sampling} uses TCBP
model training at the non-default physical error rates.},
nearly every data point in this paper was collected using a single set of default
parameters, for both TCBP model creation and JOBS searches.
Given the limited number of codes studied in this paper, we expect that many optimizations for speed or
accuracy on new codes remain for further studies.
The first open question is whether the accuracy gap to Tesseract at low physical error
rates, where TCBP-JOBS fails on some four-fault configurations (Section~\ref{sec:importance-sampling}),
can be closed by changing how the TCBP model is trained or by increasing the JOBS search effort.
Further open questions include how many BP iterations TCBP needs, how min-sum BP compares
with sum-product BP as its front end, how best to train and initialize the TCBP model,
and how to choose JOBS search parameters for a given code.

Additionally, integrating
the decoder with a quantum control system and evaluating complete
reaction times are the next steps toward establishing its suitability for real-time
fault-tolerant computation.
Such studies should evaluate sustained throughput and latency tails against the
arrival rates and feed-forward deadlines of specific hardware schedules.
Another direction is to explore alternatives to the current
$\lceil\alpha K^*\rceil$ search-window rule, assessing how window selection changes
the tradeoff between search cost and logical error rate.

\section*{Acknowledgments}

We acknowledge the use of AI-assisted tools (Anthropic and OpenAI models) in the preparation of this
manuscript. The authors reviewed and validated all resulting technical content
and retain responsibility for the scientific claims and final manuscript.

%% file: inputs/A1jobs-vs-tesseract-per-basis.tex
\providecommand{\todo}[1]{\IB{TODO: #1}}


\begin{table}[p]
\centering
\caption{Per-basis version of the comparison of TCBP-JOBS with sparsified short-beam Tesseract, one row per
circuit: bivariate-bicycle codes. Shots is the prefix decoded by both. Fails are counted on
that prefix. T-only and J-only are the shots that fail under one decoder only. Ratio is
Tesseract LER over JOBS LER. $p_{\mathrm{McN}}$ is the exact two-sided McNemar test on the
discordant shots.}
\label{tab:jobs-vs-tesseract-per-basis-bb}
\small
\setlength{\tabcolsep}{3pt}
\begin{tabular}{llcrrrrrr}
\toprule
code & $p$ & basis & shots & fails T & fails J & T-only / J-only & ratio & $p_{\mathrm{McN}}$ \\
\midrule
$[[72,12,6]]$ & 0.001 & $X$ & 55k & 33 & 35 & 8 / 10 & 0.9 & 0.81 \\
$[[72,12,6]]$ & 0.001 & $Z$ & 63k & 26 & 31 & 6 / 11 & 0.8 & 0.33 \\
$[[72,12,6]]$ & 0.002 & $X$ & 6.8k & 44 & 38 & 11 / 5 & 1.2 & 0.21 \\
$[[72,12,6]]$ & 0.002 & $Z$ & 3.0k & 25 & 22 & 6 / 3 & 1.1 & 0.51 \\
$[[72,12,6]]$ & 0.004 & $X$ & 437 & 72 & 70 & 15 / 13 & 1.0 & 0.85 \\
$[[72,12,6]]$ & 0.004 & $Z$ & 489 & 87 & 75 & 25 / 13 & 1.2 & 0.07 \\
\midrule
$[[90,8,10]]$ & 0.001 & $X$ & 19.7M & 53 & 57 & 10 / 14 & 0.9 & 0.54 \\
$[[90,8,10]]$ & 0.001 & $Z$ & 7.0M & 32 & 31 & 7 / 6 & 1.0 & 1.00 \\
$[[90,8,10]]$ & 0.002 & $X$ & 101k & 25 & 14 & 13 / 2 & 1.8 & $7 \times 10^{-3}$ \\
$[[90,8,10]]$ & 0.002 & $Z$ & 77k & 25 & 18 & 10 / 3 & 1.4 & 0.09 \\
$[[90,8,10]]$ & 0.004 & $X$ & 332 & 49 & 19 & 34 / 4 & 2.6 & $6 \times 10^{-7}$ \\
$[[90,8,10]]$ & 0.004 & $Z$ & 487 & 54 & 32 & 31 / 9 & 1.7 & $7 \times 10^{-4}$ \\
\midrule
$[[108,8,10]]$ & 0.001 & $X$ & 68.0M & 41 & 55 & 10 / 24 & 0.7 & 0.02 \\
$[[108,8,10]]$ & 0.001 & $Z$ & 42.6M & 38 & 43 & 3 / 8 & 0.9 & 0.23 \\
$[[108,8,10]]$ & 0.002 & $X$ & 225k & 25 & 11 & 18 / 4 & 2.3 & $4 \times 10^{-3}$ \\
$[[108,8,10]]$ & 0.002 & $Z$ & 176k & 26 & 10 & 17 / 1 & 2.6 & $1 \times 10^{-4}$ \\
$[[108,8,10]]$ & 0.004 & $X$ & 393 & 59 & 11 & 49 / 1 & 5.4 & $<\!10^{-10}$ \\
$[[108,8,10]]$ & 0.004 & $Z$ & 427 & 56 & 20 & 42 / 6 & 2.8 & $1 \times 10^{-7}$ \\
\midrule
$[[144,12,12]]$ & 0.001 & $X$ & 133.7M & 26 & 23 & 11 / 8 & 1.1 & 0.65 \\
$[[144,12,12]]$ & 0.001 & $Z$ & 88.0M & 27 & 32 & 6 / 11 & 0.8 & 0.33 \\
$[[144,12,12]]$ & 0.002 & $X$ & 196k & 25 & 5 & 21 / 1 & 5.0 & $1 \times 10^{-5}$ \\
$[[144,12,12]]$ & 0.002 & $Z$ & 159k & 25 & 6 & 19 / 0 & 4.2 & $4 \times 10^{-6}$ \\
$[[144,12,12]]$ & 0.004 & $X$ & 226 & 65 & 9 & 56 / 0 & 7.2 & $<\!10^{-10}$ \\
$[[144,12,12]]$ & 0.004 & $Z$ & 226 & 56 & 17 & 41 / 2 & 3.3 & $2 \times 10^{-10}$ \\
\bottomrule
\end{tabular}
\end{table}

\begin{table}[p]
\centering
\caption{Per-basis comparison, continued: color codes. Columns as in
Table~\ref{tab:jobs-vs-tesseract-per-basis-bb}.}
\label{tab:jobs-vs-tesseract-per-basis-color}
\small
\setlength{\tabcolsep}{3pt}
\begin{tabular}{llcrrrrrr}
\toprule
code & $p$ & basis & shots & fails T & fails J & T-only / J-only & ratio & $p_{\mathrm{McN}}$ \\
\midrule
color $d{=}7$ & 0.001 & $X$ & 112k & 28 & 30 & 2 / 4 & 0.9 & 0.69 \\
color $d{=}7$ & 0.001 & $Z$ & 88k & 27 & 28 & 4 / 5 & 1.0 & 1.00 \\
color $d{=}7$ & 0.002 & $X$ & 7.0k & 57 & 50 & 10 / 3 & 1.1 & 0.09 \\
color $d{=}7$ & 0.002 & $Z$ & 7.0k & 47 & 44 & 9 / 6 & 1.1 & 0.61 \\
color $d{=}7$ & 0.004 & $X$ & 696 & 54 & 48 & 14 / 8 & 1.1 & 0.29 \\
color $d{=}7$ & 0.004 & $Z$ & 712 & 78 & 86 & 8 / 16 & 0.9 & 0.15 \\
\midrule
color $d{=}9$ & 0.001 & $X$ & 758k & 27 & 28 & 9 / 10 & 1.0 & 1.00 \\
color $d{=}9$ & 0.001 & $Z$ & 463k & 25 & 26 & 3 / 4 & 1.0 & 1.00 \\
color $d{=}9$ & 0.002 & $X$ & 7.0k & 26 & 25 & 6 / 5 & 1.0 & 1.00 \\
color $d{=}9$ & 0.002 & $Z$ & 9.8k & 27 & 26 & 7 / 6 & 1.0 & 1.00 \\
color $d{=}9$ & 0.004 & $X$ & 446 & 46 & 51 & 22 / 27 & 0.9 & 0.57 \\
color $d{=}9$ & 0.004 & $Z$ & 373 & 49 & 48 & 21 / 20 & 1.0 & 1.00 \\
\midrule
color $d{=}11$ & 0.001 & $X$ & 8.0M & 48 & 57 & 17 / 26 & 0.8 & 0.22 \\
color $d{=}11$ & 0.001 & $Z$ & 7.5M & 86 & 78 & 31 / 23 & 1.1 & 0.34 \\
color $d{=}11$ & 0.002 & $X$ & 12k & 26 & 12 & 17 / 3 & 2.2 & $3 \times 10^{-3}$ \\
color $d{=}11$ & 0.002 & $Z$ & 13k & 26 & 24 & 11 / 9 & 1.1 & 0.82 \\
color $d{=}11$ & 0.004 & $X$ & 292 & 50 & 32 & 38 / 20 & 1.6 & 0.02 \\
color $d{=}11$ & 0.004 & $Z$ & 312 & 61 & 42 & 44 / 25 & 1.5 & 0.03 \\
\midrule
color $d{=}13$ & 0.001 & $X$ & 16.8M & 53 & 30 & 41 / 18 & 1.8 & $4 \times 10^{-3}$ \\
color $d{=}13$ & 0.001 & $Z$ & 14.0M & 61 & 30 & 44 / 13 & 2.0 & $5 \times 10^{-5}$ \\
color $d{=}13$ & 0.002 & $X$ & 9.0k & 26 & 4 & 24 / 2 & 6.5 & $1 \times 10^{-5}$ \\
color $d{=}13$ & 0.002 & $Z$ & 11k & 27 & 8 & 24 / 5 & 3.4 & $5 \times 10^{-4}$ \\
color $d{=}13$ & 0.004 & $X$ & 228 & 74 & 35 & 64 / 25 & 2.1 & $4 \times 10^{-5}$ \\
color $d{=}13$ & 0.004 & $Z$ & 223 & 98 & 53 & 67 / 22 & 1.8 & $2 \times 10^{-6}$ \\
\bottomrule
\end{tabular}
\end{table}

\begin{table}[p]
\centering
\caption{Per-basis comparison, continued: rotated surface codes. Columns as in
Table~\ref{tab:jobs-vs-tesseract-per-basis-bb}.}
\label{tab:jobs-vs-tesseract-per-basis-surface}
\small
\setlength{\tabcolsep}{3pt}
\begin{tabular}{llcrrrrrr}
\toprule
code & $p$ & basis & shots & fails T & fails J & T-only / J-only & ratio & $p_{\mathrm{McN}}$ \\
\midrule
surface $d{=}7$ & 0.001 & $X$ & 271k & 25 & 24 & 1 / 0 & 1.0 & 1.00 \\
surface $d{=}7$ & 0.001 & $Z$ & 305k & 25 & 22 & 3 / 0 & 1.1 & 0.25 \\
surface $d{=}7$ & 0.002 & $X$ & 15k & 26 & 24 & 2 / 0 & 1.1 & 0.50 \\
surface $d{=}7$ & 0.002 & $Z$ & 15k & 32 & 30 & 2 / 0 & 1.1 & 0.50 \\
surface $d{=}7$ & 0.004 & $X$ & 1.0k & 31 & 30 & 3 / 2 & 1.0 & 1.00 \\
surface $d{=}7$ & 0.004 & $Z$ & 1.0k & 27 & 23 & 6 / 2 & 1.2 & 0.29 \\
\midrule
surface $d{=}9$ & 0.001 & $X$ & 8.0M & 83 & 83 & 2 / 2 & 1.0 & 1.00 \\
surface $d{=}9$ & 0.001 & $Z$ & 5.2M & 58 & 57 & 2 / 1 & 1.0 & 1.00 \\
surface $d{=}9$ & 0.002 & $X$ & 54k & 26 & 25 & 2 / 1 & 1.0 & 1.00 \\
surface $d{=}9$ & 0.002 & $Z$ & 61k & 29 & 26 & 3 / 0 & 1.1 & 0.25 \\
surface $d{=}9$ & 0.004 & $X$ & 1.0k & 28 & 25 & 8 / 5 & 1.1 & 0.58 \\
surface $d{=}9$ & 0.004 & $Z$ & 1.0k & 29 & 25 & 7 / 3 & 1.2 & 0.34 \\
\midrule
surface $d{=}11$ & 0.001 & $X$ & 29.2M & 42 & 36 & 7 / 1 & 1.2 & 0.07 \\
surface $d{=}11$ & 0.001 & $Z$ & 30.0M & 40 & 33 & 11 / 4 & 1.2 & 0.12 \\
surface $d{=}11$ & 0.002 & $X$ & 84k & 26 & 16 & 11 / 1 & 1.6 & $6 \times 10^{-3}$ \\
surface $d{=}11$ & 0.002 & $Z$ & 185k & 27 & 23 & 9 / 5 & 1.2 & 0.42 \\
surface $d{=}11$ & 0.004 & $X$ & 1.0k & 32 & 22 & 17 / 7 & 1.5 & 0.06 \\
surface $d{=}11$ & 0.004 & $Z$ & 1.0k & 31 & 12 & 21 / 2 & 2.6 & $7 \times 10^{-5}$ \\
\midrule
surface $d{=}13$ & 0.001 & $X$ & 77.0M & 33 & 12 & 21 / 0 & 2.8 & $1 \times 10^{-6}$ \\
surface $d{=}13$ & 0.001 & $Z$ & 105.5M & 26 & 11 & 16 / 1 & 2.4 & $3 \times 10^{-4}$ \\
surface $d{=}13$ & 0.002 & $X$ & 157k & 26 & 8 & 18 / 0 & 3.2 & $8 \times 10^{-6}$ \\
surface $d{=}13$ & 0.002 & $Z$ & 101k & 25 & 9 & 17 / 1 & 2.8 & $1 \times 10^{-4}$ \\
surface $d{=}13$ & 0.004 & $X$ & 843 & 34 & 13 & 27 / 6 & 2.6 & $3 \times 10^{-4}$ \\
surface $d{=}13$ & 0.004 & $Z$ & 949 & 40 & 13 & 31 / 4 & 3.1 & $3 \times 10^{-6}$ \\
\bottomrule
\end{tabular}
\end{table}

%% file: inputs/A2serial_latency.tex
\section{Serial decoding latency on the bivariate bicycle codes}
\label{app:serial_latency}

This appendix supports the serial-latency comparison in the main text but is focused on a
comparison to the unsparsified Tesseract long-beam preset rather than the sparsified long-beam preset.
Latency means one shot decoded at a time
with no batching: Tesseract as a single-threaded process, TCBP-JOBS as single-shot using as much
of the GPU as it is configured to use.
Both decoders decoded the same 2{,}000 Stim-sampled shots per code
(SI1000, $p=0.2\%$, Z basis, Tesseract benchmark circuits). Tesseract~\cite{beni2025tesseract} ran its long-beam preset without sparsification
(\texttt{--beam 20 --beam-climbing --num-det-orders 21 --pqlimit $10^6$ --no-revisit-dets --det-order-index}, commit
54569b4) on Grace (Neoverse-V2) cores of a GB200 node. TCBP-JOBS ran on one GB200 GPU of the same node.

Tesseract's decode time is sensitive to system load: with one process per core on all cores the mean per shot is 1.1 to 3.8
times higher than on an otherwise idle machine. However, obtaining data with only one process per machine is prohibitively 
computationally expensive. We first collect long-beam latency data using all cores and then we
obtain the \emph{quiet-core} mean, by re-decoding a random subset of the same shots alone on an idle machine
(30 shots per code; 160 for $[[144,12,12]]$), forming the per-shot loaded/quiet ratio, and dividing the 2{,}000-shot
loaded mean by it. The published mean for the same circuit and flags~\cite{tesseract-pr254} is shown for reference;
its hardware configuration (CPU architecture and number of cores per machine) is not specified, so a
direct comparison cannot be made.

\begin{table}[h]
\centering
\small
\caption{Serial per-shot decoding time, same 2{,}000 shots per code, SI1000 $p=0.2\%$, Z basis. Tesseract ran its
long-beam preset without sparsification; its times are quiet-core means (see text); TCBP-JOBS times are single-shot \texttt{decode()} on one GB200. Speedup is the ratio of means. PR254 refers to \cite{tesseract-pr254}.}
\label{tab:serial_latency}
\begin{tabular}{l r r r r r r}
\toprule
 & \multicolumn{2}{c}{Tesseract long beam, 1 Grace core} & \multicolumn{2}{c}{TCBP-JOBS, 1 GB200} & \multicolumn{2}{c}{speedup} \\
\cmidrule(lr){2-3} \cmidrule(lr){4-5} \cmidrule(lr){6-7}
code & quiet-core mean & PR254 mean & mean & p99 & vs.\ quiet core & vs.\ PR254 \\
\midrule
$[[72,12,6]]$   & 7.5 s  & 13.0 s & 1.23 ms & 2.3 ms  & 6{,}100  & 10{,}600 \\
$[[90,8,10]]$   & 19.4 s & 35.2 s & 1.91 ms & 5.0 ms  & 10{,}100 & 18{,}400 \\
$[[108,8,10]]$  & 31.5 s & 44.7 s & 2.36 ms & 7.7 ms  & 13{,}300 & 18{,}900 \\
$[[144,12,12]]$ & 53.5 s & 70.8 s & 4.09 ms & 16.0 ms & 13{,}100 & 17{,}300 \\
\bottomrule
\end{tabular}
\end{table}

As Table~\ref{tab:serial_latency} shows, the speedup measured on our own GB200 hardware is 6{,}000 to 13{,}000 times; against PR254's published means it is 11{,}000 to
19{,}000 times. TCBP-JOBS has a short tail: its p99 is within 4 times its mean on every code. The two decoders' accuracy at these
settings is compared in Appendix~\ref{app:batch_throughput} on 0.5 to 2.6 million shots per code.

%% file: inputs/A3batch_throughput.tex
\section{Batched decoding: the Tesseract benchmark at $p=0.2\%$}
\label{app:batch_throughput}

This section complements the data in Section~\ref{sec:jobs-vs-tesseract} and is focused on unsparsified long-beam
comparisons whereas the main paper focused on sparsified comparisons (both long-beam and short-beam).

In a research setting the quantity of interest is the time to decode a large number of shots, so we compare throughput
rather than latency. We decoded every circuit of the Tesseract~\cite{beni2025tesseract} SI1000 benchmark at $p=0.2\%$ (surface, color and
bivariate bicycle codes, X and Z bases, 28 circuits) with TCBP-JOBS, using for each circuit and basis exactly the
number of shots that were published with the unsparsified long-beam preset in 
results~\cite{tesseract-pr254}, 24.2 million shots in all. TCBP-JOBS ran with preset
\texttt{tcbp-jobs-sc} (10 sum-product BP iterations, JOBS with the TCBP ordering, 16 draws), which differs from the
configuration of Table~\ref{tab:jobs-settings} only in that JOBS also searches the BP-converged shots and returns the
cheaper of BP's correction and its own. It ran on one GB200 GPU with
\texttt{decode\_batch} calls of 16{,}384 shots; time is the wall time of the decode calls, including this additional search. Tesseract's
time is its own per-shot decode time summed over worker threads. Both decoders used the circuits' own detector error models. Table~\ref{tab:batch_accuracy} gives the
accuracy and the per-shot time on each side with X and Z pooled; Table~\ref{tab:batch_walltime} converts the totals
into wall time, assuming Tesseract's thread-hours spread perfectly over a 64-core server.

\begin{table*}[t]
\centering
\footnotesize
\setlength{\tabcolsep}{4pt}
\caption{Tesseract long beam without sparsification~\cite{tesseract-pr254} versus TCBP-JOBS on the Tesseract
benchmark at $p=0.2\%$, X and Z bases pooled, identical shot counts per circuit and basis. LER is the logical error
rate per shot (any observable wrong; Tesseract low-confidence shots counted as errors). Ratios are Tesseract over
TCBP-JOBS; the interval is the 95\% Clopper--Pearson bound on each LER, combined conservatively.
Totals over the 24.2 million shots: Tesseract 130{,}257 thread-hours, TCBP-JOBS 44.8 GPU-minutes.}
\label{tab:batch_accuracy}
\begin{tabular}{l r r r r r r r r r}
\toprule
 & & \multicolumn{3}{c}{Tesseract, 1 CPU thread} & \multicolumn{3}{c}{TCBP-JOBS, 1 GB200, batched} & \multicolumn{2}{c}{ratio T/J} \\
\cmidrule(lr){3-5} \cmidrule(lr){6-8} \cmidrule(lr){9-10}
code & shots & fails & LER & s/shot & fails & LER & $\mu$s/shot & LER [95\% CI] & time \\
\midrule
surface d3 & 2.60M & 29{,}491 & $1.13\times10^{-2}$ & 0.000202 & 29{,}713 & $1.14\times10^{-2}$ & 5 & 0.99 [1.0, 1.0] & 41 \\
surface d5 & 0.82M & 3{,}806 & $4.63\times10^{-3}$ & 0.00425 & 3{,}735 & $4.54\times10^{-3}$ & 12 & 1.02 [1.0, 1.1] & 343 \\
surface d7 & 2.16M & 3{,}269 & $1.52\times10^{-3}$ & 0.0299 & 3{,}407 & $1.58\times10^{-3}$ & 20 & 0.96 [0.9, 1.0] & 1{,}471 \\
surface d9 & 2.60M & 1{,}228 & $4.73\times10^{-4}$ & 0.123 & 1{,}249 & $4.81\times10^{-4}$ & 36 & 0.98 [0.9, 1.1] & 3{,}408 \\
surface d11 & 2.60M & 390 & $1.50\times10^{-4}$ & 0.384 & 388 & $1.49\times10^{-4}$ & 72 & 1.01 [0.8, 1.2] & 5{,}350 \\
\addlinespace
color d3 & 0.58M & 12{,}269 & $2.13\times10^{-2}$ & 0.00137 & 12{,}118 & $2.10\times10^{-2}$ & 5 & 1.01 [1.0, 1.0] & 266 \\
color d5 & 1.04M & 14{,}073 & $1.35\times10^{-2}$ & 0.149 & 14{,}089 & $1.35\times10^{-2}$ & 15 & 1.00 [1.0, 1.0] & 10{,}145 \\
color d7 & 0.60M & 3{,}777 & $6.27\times10^{-3}$ & 2.89 & 3{,}941 & $6.54\times10^{-3}$ & 34 & 0.96 [0.9, 1.0] & 85{,}594 \\
color d9 & 1.10M & 2{,}762 & $2.51\times10^{-3}$ & 11.3 & 3{,}079 & $2.80\times10^{-3}$ & 80 & 0.90 [0.8, 1.0] & 141{,}348 \\
color d11 & 1.83M & 2{,}577 & $1.41\times10^{-3}$ & 28.2 & 2{,}060 & $1.13\times10^{-3}$ & 204 & 1.25 [1.2, 1.4] & 138{,}377 \\
\addlinespace
$[[72,12,6]]$ & 0.54M & 3{,}304 & $6.09\times10^{-3}$ & 13.1 & 3{,}407 & $6.28\times10^{-3}$ & 55 & 0.97 [0.9, 1.0] & 239{,}983 \\
$[[90,8,10]]$ & 2.60M & 577 & $2.22\times10^{-4}$ & 35.3 & 566 & $2.18\times10^{-4}$ & 135 & 1.02 [0.9, 1.2] & 261{,}448 \\
$[[108,8,10]]$ & 2.60M & 239 & $9.19\times10^{-5}$ & 46.0 & 197 & $7.58\times10^{-5}$ & 178 & 1.21 [0.9, 1.6] & 257{,}978 \\
$[[144,12,12]]$ & 2.58M & 144 & $5.58\times10^{-5}$ & 71.0 & 92 & $3.57\times10^{-5}$ & 388 & 1.57 [1.1, 2.3] & 183{,}119 \\
\midrule
geometric mean & & & & & & & & 1.05 & 13{,}242 \\
\bottomrule
\end{tabular}
\end{table*}

\begin{table}[h]
\centering
\small
\caption{Wall time to decode the benchmark of Table~\ref{tab:batch_accuracy}: Tesseract's thread-hours spread perfectly
over a 64-core server versus one GB200 GPU. Ratios are of summed times.}
\label{tab:batch_walltime}
\begin{tabular}{l r r r r}
\toprule
circuits & shots & Tesseract, 64 cores & TCBP-JOBS, 1 GB200 & ratio \\
\midrule
surface codes (10)          & 10.8 M & 6.0 h      & 5.8 min & 62 \\
color codes (10)           & 5.1 M  & 11.9 days  & 8.3 min & 2{,}060 \\
bivariate bicycle codes (8) & 8.3 M  & 10.4 weeks & 31 min  & 3{,}400 \\
\midrule
all 28                      & 24.2 M & 12.1 weeks & 45 min  & 2{,}720 \\
\bottomrule
\end{tabular}
\end{table}

TCBP-JOBS matches Tesseract's accuracy over the benchmark (geometric-mean LER ratio 1.05, with 1.25 at color $d=11$ and
1.57 at $[[144,12,12]]$ in its favor and 0.90 at color $d=9$ against it) while decoding the full set in 45 minutes on
one GPU. The same work is 130{,}000 thread-hours for Tesseract, or about 12 weeks on a 64-core server; on the bivariate
bicycle codes alone it is 10 weeks against half an hour.

%% file: inputs/A4splitting.tex
\section{Rare-event splitting}
\label{app:splitting}

This appendix describes the logical error rate estimator of Section~\ref{sec:importance-sampling}. We summarize the splitting method of Bravyi and Vargo~\cite{bravyi2013simulation}, and then discuss several modifications we made to apply it to TCBP-JOBS on the gross code. 

We use the notation of Section~\ref{sec:background}. The detector error model (DEM) of the 12-round $Z$ memory circuit of the $[[144,12,12]]$ code has $m = 1{,}728$ detectors and $n = 67{,}752$ faults, with check matrix $H$ and logical matrix $L$ for its 12 observables. A fault configuration is a vector $F \in \{0,1\}^n$ whose entry $F_j$ is 1 if fault $j$ occurred, and its weight is the number of faults it contains. The faults are the independent error mechanisms in the DEM derived using an SI1000 circuit noise model \cite{gidney2021faulttolerant}. This model has a single physical error rate $p$ that scales all noise parameters, with $q_j(p)$ denoting the resulting probability of fault $j$.  Faults occur independently, so a configuration has probability
\begin{equation}
  \Pr\nolimits_p(F) = \prod_{j=1}^{n} q_j(p)^{F_j} \bigl(1 - q_j(p)\bigr)^{1 - F_j}.
  \label{eq:config-prob}
\end{equation}
We distinguish these true rates $q_j(p)$, which generate the faults, from the prior $p_j$ that the decoder is given. The syndrome of $F$ is $\sigma = HF$. The decoder maps $\sigma$ to a prediction of the observable flips, and it fails on $F$ if that prediction differs from $LF$ in any of the 12 observables. We write $\calB$ for the set of configurations on which the decoder fails. The logical error rate per 12 rounds is then $P_{\mathrm{L}}(p) = \sum_{F \in \calB} \Pr_p(F)$, and the distribution of failures at $p$ is $\pi_p(F) = \Pr_p(F) / P_{\mathrm{L}}(p)$ for $F \in \calB$ and zero otherwise.

\subsection{Bravyi and Vargo splitting method}
Direct Monte Carlo sampling draws configurations from $\Pr_p$ and counts how many fall in $\calB$, which needs of order $1 / P_{\mathrm{L}}(p)$ shots. Splitting avoids this by considering only failing configurations $F\in\calB$. Choose a decreasing sequence of physical error rates $p_0 > p_1 > \dots > p_K$, also called \textit{levels}, with $p_0$ large enough that $P_{\mathrm{L}}(p_0)$ can be measured directly. Then
\begin{equation}
  P_{\mathrm{L}}(p_K) = P_{\mathrm{L}}(p_0) \prod_{k=0}^{K-1} \frac{P_{\mathrm{L}}(p_{k+1})}{P_{\mathrm{L}}(p_k)},
  \qquad
  \frac{P_{\mathrm{L}}(p_{k+1})}{P_{\mathrm{L}}(p_k)} = \EE_{F \sim \pi_{p_k}}\bigl[ r_k(F) \bigr],
  \qquad
  r_k(F) = \frac{\Pr_{p_{k+1}}(F)}{\Pr_{p_k}(F)}.
  \label{eq:splitting}
\end{equation}
The first equality is a telescoping product. The second follows from writing both logical error rates as sums over the same set $\calB$. Each ratio is the average of the reweighting factor $r_k$ over failures at the higher level. The factor $r_k(F)$ is computed from Eq.~\eqref{eq:config-prob}, without calling the decoder. If neighboring levels are close, $r_k$ varies little over the failures at $p_k$ and a modest number of samples gives each ratio accurately, although the product of all ratios may be many orders of magnitude.

The samples of $\pi_{p_k}$ come from a Markov chain. Bravyi and Vargo use a Metropolis chain that proposes flipping one uniformly chosen fault, accepts the proposal with the usual probability $\min[1, \Pr_{p_k}(F') / \Pr_{p_k}(F)]$, and rejects it whenever the proposed configuration $F'$ is not in $\calB$, which takes one decoder call to decide. They estimate each ratio from the chains at both of its levels with an acceptance-ratio estimator, which combines samples of $\pi_{p_k}$ and $\pi_{p_{k+1}}$ and has a lower variance than the one-sided average in Eq.~\eqref{eq:splitting}. Their levels are spaced by $p_{k+1} = p_k\, 2^{-1/\sqrt{w_k}}$, where $w_k = \max(d/2,\, p_k N)$ for a code of distance $d$ with $N$ fault locations is roughly the weight of a typical failure at $p_k$. Each chain starts from an uncorrectable error constructed by hand, and $P_{\mathrm{L}}(p_0)$ comes from an asymptotic formula or from Monte Carlo sampling. Their simulations are of the surface code decoded by minimum-weight matching. Equation~\eqref{eq:splitting} itself assumes nothing about the code or the decoder beyond the requirement that the decoder must be a fixed function of the syndrome, so that $\calB$ is a well-defined set that is the same at every level.

\subsection{Adapting splitting to TCBP-JOBS on the gross code}
We made several changes to the splitting method to optimize its performance for TCBP-JOBS on the gross code. We kept Eq.~\eqref{eq:splitting} and changed how the failures are sampled, how each ratio is estimated, and how a run starts. These changes address both the size of the DEM and the increased throughput of the GPU decoder when using batched execution. 

\subsubsection{Chain move}
Evolving the chains by flipping one fault per step leads to slow exploration. A failing configuration at $p_0 = 0.004$ contains about 130 of the 67{,}752 faults, most of which do not contribute to the failure, and a uniformly chosen flip rarely changes a significant fault. We use the partial-resampling move of Ref.~\cite{mullan2026improved} instead. At level $k$, each step selects every fault independently with probability $1/4$ and redraws each selected fault from its own rate $q_j(p_k)$, leaving the others as they are. Redrawing a subset of independent coordinates from their own distribution leaves $\Pr_{p_k}$ unchanged, so there is no Metropolis acceptance probability to compute, and the proposal is accepted only if the new configuration remains in $\calB$. Each step therefore costs one decoder call per chain. The fraction of accepted proposals is between 37\% and 47\% at $p = 0.004$, depending on the decoder of the run, and between 8\% and 15\% below $p = 0.002$.

\subsubsection{Population and resampling}
A run evolves 128 chains together and sends their 128 syndromes to the decoder as one batch at every step. Multiple chains are required because single chains rarely move between failures caused by different small sets of faults. Transitions between failures caused by different small sets of faults can require rare intermediate configurations, making such transitions infrequent at low $p$. At the end of level $k$, we resample the 128 final configurations with replacement, each with probability proportional to the average of $r_k$ along its chain, and the 128 copies are the starting configurations of level $k+1$. Types of failure that become relatively more likely at the lower error rate are thereby duplicated and the others are thinned. This reduces bias but comes at a loss of diversity. In the runs to $p = 10^{-4}$, the 128 chains at the last level descend from a median of 22 of the 128 starting configurations. Runs to higher targets pass through fewer levels and retain more, with a median of 33 at $p = 10^{-3}$.

\subsubsection{Ratio estimate}
Each ratio in Eq.~\eqref{eq:splitting} is estimated as the average of $r_k(F)$ over the second half of the steps of every chain at level $k$. The first half is discarded to allow the chains to relax to $\pi_{p_k}$. We did not use the acceptance-ratio estimator given in Ref.~\cite{bravyi2013simulation}.

\subsubsection{Chain evolution}
Let $\mu(p)=\sum_j q_j(p)$ be the expected number of faults in a shot, which is 126 at $p=0.004$ and 3.2 at $p=10^{-4}$. We space the levels so that $|\ln p_{k+1}-\ln p_k|\le 2/\sqrt{\mu(p_k)}$, which has the same form as the rule of Ref.~\cite{bravyi2013simulation}, with steps about three times as wide.

For each target physical error rate $p$, we use a decoder trained at that target and hold it fixed throughout the splitting run. Each run begins with 20,000 Monte Carlo shots at $p_0=0.004$, using that run's decoder to estimate $P_{\mathrm{L}}(p_0)$ and obtain failing configurations to initialize the 128 chains.

The level-spacing rule gives 12 steps for the full interval from $p_0=0.004$ to $p=10^{-4}$; runs targeting higher $p$ terminate earlier. Every chain takes 1,000 steps at each of the first two levels and 2,000 steps at each subsequent level used to estimate a ratio. For the $p=10^{-4}$ endpoint, this requires approximately $2.8\times10^6$ decoder calls per run, or $2.2\times10^7$ calls across eight independent runs.

\subsubsection{Uncertainty}
We performed eight independent MCMC runs for each target $p$, each with a separate random seed and Monte Carlo sample at $p_0$. We assess statistical variability across independent runs because resampling gives chains within a run shared ancestors. Figure~\ref{fig:splitting-ler} shows the median and full range of the logical error rate estimates. This range is not a confidence interval and does not capture a bias shared by all runs, such as that caused by incomplete exploration of the failure set.

\subsubsection{Limits}
The spread across runs cannot indicate any bias that may be shared by all runs. If every chain misses a class of failures, the runs can agree while underestimating the logical error rate. Monte Carlo using same decoder validates the splitting estimates at five error rates from $p = 0.003$ to $p = 0.001$. The ratio of the splitting estimate to the Monte Carlo estimate lies between 0.79 and 0.98, each within about two standard errors of 1, but all five lie below 1, with a weighted mean of $0.92 \pm 0.04$. This check does not establish that the chains explore all relevant failures at lower $p$.

We also computed the Gelman--Rubin statistic $\hat R$ for $\ln r_k$ at each level \cite{gelman1992inference}. This statistic compares variation between chains with variation within chains and approaches 1 as the chains converge to the same distribution. The largest $\hat R$ across levels is 1.2 in a typical run and 1.3 in the worst run, indicating that the chains are not completely mixing. As discussed before, this is unsurprising, as transitions between distinct failing sectors are suppressed.